\documentstyle[preprint,eqsecnum,prb,aps,tighten,epsf]{revtex}
\begin{document}
\preprint{UBCTP-95-006}
\draft
\title{Theory of Nuclear Magnetic Relaxation in Haldane Gap Antiferromagnets}
 \author{Jacob Sagi$^a$ and Ian Affleck$^b$}
\address{$^{a}$Physics Department} \address{University of British
Columbia, Vancouver, BC, V6T 1Z1, Canada}
\address{$^{b}$Canadian Institute for Advanced
Research and Physics Department} \address{University of British
Columbia, Vancouver, BC, V6T 1Z1, Canada} \date{\today} \maketitle
\begin{abstract}
A Theory of Nuclear Magnetic Resonance (NMR) is developed for integer-spin,
one-dimensional antiferromagnets, which exhibit the Haldane gap.  We consider
free boson, free fermion and non-linear $\sigma$-model approaches, all of which
give similar results.  Detailed anisotropy and magnetic field dependence is
calculated and compared with experiment.
\end{abstract}
\section{Introduction}
The one-dimensional Heisenberg antiferromagnet was argued to have a gap for
integer (but not half-integer) spin by Haldane \cite{hal1},
in 1983.  Since that time a
considerable amount of theoretical and experimental work has been done on this
system. Many properties of this sytem have been calculated using relativistic
quantum field theories in (1+1) space-time dimensions.  While the non-linear
$\sigma$-model is probably the most firmly established model for this system,
only a few of its features can be calculated exactly, so other
more approximate models have been used extensively, notably free boson and free
fermion models.  All three models are generally expected to give qualitatively
similar results. These models have been used to analyze neutron-scattering,
susceptibility, electron spin resonance and other experiments (see
[\onlinecite{aff1,tsev,Aff,wes,mit,aff,joli}]
and references therein).  Our purpose
here is to develop this theoretical framework to study nuclear magnetic
resonance and to compare the theory to published experiments.  Some of our
results were obtained independently by Jolicoeur and Golinelli \cite{joli}.

We consider the general Hamiltonian for $S=1$:
 \begin{equation}
H=\sum_i\{J{\vec S}_{i}\cdot{\vec S}_{i+1}+D(S_i^z)^2
+E\left[(S_i^x)^2-(S_i^y)^2\right] - \mu_B \vec{H} \cdot {\bf G}
\cdot \vec{S}_i
\}.
\label{eq:h}
\end{equation}
For sufficiently small $D/J$ and $E/J$, the model is still in the Haldane gap
phase.  In the pure Heisenberg model ($D=E=0$) the lowest excited state is a
triplet, at energy $\Delta \approx .4105 J$ above the singlet groundstate.
The $D$ term splits the triplet into a singlet and doublet, and the
$E$ term further splits
the doublet. A uniform magnetic field lowers the energy of one element of this
triplet; it vanishes at a critical field, $H_c$.  If the Hamiltonian is
invariant
under rotations about the field direction (eg. $E=0$ and the field is in the
$z$
direction), then there is quasi-long-range antiferromagnetic order above $H_c$.
If the Hamiltonian does not have this invariance, there is true long-range
order
above $H_c$.
The gyromagnetic tensor, $\bf{G}$, is often assumed diagonal
although we consider more general situations (see Appendix \ref{appC}.)

The NMR  rate, $1/T_1$, is given by:
\begin{eqnarray}
1/T_1 &\propto& \sum_{i,j}
\int_{-\infty}^\infty
dte^{-i\omega_Nt}A^{bi}A^{jb}<\{ S^i(0,t),S^j(0,0)\} >\nonumber
\\ &\propto &\sum_{i,j}
A^{bi}A^{jb} \sum_{n,m}e^{-E_n/T}\left(1+e^{\omega_N/T}\right)<n|S^i(0)|m>
<m|S^j(0)|n> \nonumber \\ && 2\pi \delta
(E_n-E_m-\omega_N). \label{1/T_1}
\end{eqnarray}

Here $A^{jb}$
is the hyperfine coupling between the nucleus  and the
Ni spin.  $b$ is the direction of the RF field. In NENP,
we will generally assume that
$A^{ab}(q)$ is not strongly $q$-dependent and is independent of $b$ after
averaging over the various nuclei.
 $<>$ denotes a thermal average
in an applied magnetic field, $H$ (normally perpendicular to the RF field).
$\omega_N\equiv \mu_NH$ is the nuclear Larmor frequency.
 This is of order $1mK$, negligibly small compared to the
other relevant energy scales. Thus we are concerned with transitions between
states $n$ and $m$ with essentially no change in energy but arbitrary change in
wave-vector. We also can essentially set $e^{\omega_N/T} = 1$.
We will concentrate on calculating $1/T_1$ in
the low $T$ limit, $T \ll \Delta$, the
Haldane gap, where the low energy field theory approximation may be used.  In
this limit, there are contributions to $1/T_1$ from $q\approx \pi$ and
$q\approx
0$.

There are several facets to a calculation of the NMR rate.  We must consider
contributions from two different ranges of $q$.  Furthermore, the nature of the
excitations changes radically near the critical field.  As we vary the field at
fixed temperature, we may identify three different regions:
$\mu_B(H_c-H) \gg T$, $\mu_B|H-H_c|$ of O(T), and
$\mu_B(H_c-H) \ll T$. The behaviour of $1/T_1$ is quite
different depending on whether or not there is rotational
symmetry around the direction of the static field.  Finally, the presence of
defects, ie. chain ends, may lead to additional structure at temperatures well
below $\Delta$.

In this paper, we mainly focus on the regime $\mu_B(H_c-H) \gg T$, $\Delta  \gg
T$.
In the rotationally invariant case, we give an asymptotically exact expression
for
the rate which is quite independent of any detailed assumptions underlying the
field theories.  Without rotational symmetry more complicated and somewhat
model-dependent expressions are obtained.  In both cases $q\approx 0$ dominates
at low $T$.  Our result in the isotropic, zero field limit, was obtained
previously by Jolicoeur and Golinelli \cite{joli}.

We also give a partial analysis of the region where $\mu_B(H_c-H) \ll T$.
We show that the free fermion model gives an exact description of this region,
for $q\approx 0$, with or without axial symmetry. We also consider $q\approx
\pi$ in this field region, showing that it actually dominates over $q\approx
0$.

Where possible we compare our results to the experiments of Fujiwara et al.
\cite{fuji} on
Ni(C$_2$H$_8$N$_2$)$_2$NO$_2$(ClO$_4$) (NENP).   While the overall agreement is
fairly good, one puzzling discrepancy is encountered.

In Sec. II we briefly review the field theory treatment of this system and
estimate the size of the $q\approx \pi$ and $q\approx 0$ contributions to
$1/T_1$.  In particular, we review and generalize arguments that the free
fermion
model should become exact near the critical field.
Sec. III considers the contributions to $1/T_1$ below the critical field for
the isotropic and anisotropic cases. Sec. IV partially analyzes the region near
criticality. Appendix \ref{appA} contains the details of the model
calculations. Appendix \ref{appB} suggests a possible additional contribution
to
$1/T_1$ at low fields and low temperature.
We expect the presence of non-magnetic defects, which essentially
break up the system into finite length chains, to have a radical effect on the
low
energy properties of the system.  Essentially decoupled $S=1/2$ degrees of
freedom appear at each chain end. This mechanism is not in
itself sufficient to fully explain the low field experimental data.
Finally, appendix \ref{appC}
makes note of structural properties
of NENP which are expected to have effects on experiment.

\section{Overview}
At least three different field theories have been used extensively to discuss
Haldane gap antiferromagnets.  The first, and probably best founded is the
non-linear $\sigma$-model \cite{hal1}.
The antiferromagnetic order parameter, is
represented by a three-component field, $\vec \phi$, obeying a constraint,
$\vec \phi^2=1$.  The uniform magnetisation density is represented by the
conserved density, $\vec l$, of the field theory:
\begin{equation} \vec l \equiv (1/vg)\vec \phi \times \partial \vec \phi
/\partial t.\end{equation} Thus the spin operator at site $i$ is written:
\begin{equation} \vec S_i \approx s(-1)^i \vec \phi + \vec l,\end{equation}
and the Heisenberg Hamiltonian density as:
\begin{equation} {\cal H} \approx {vg\over 2}\vec l^2 + {v\over 2g}\left({d\vec
\phi \over dx}\right)^2.\end{equation}  This is expected to give the asympotic
low-energy long-distance behaviour of the Heisenberg model at large,
integer-valued spin magnitude, $s$.  The spin-wave velocity and coupling
constant
are given (up to higher order in $1/s$) by:
\begin{equation} v=2Js, g=2/s.\end{equation}  Due to strong quantum
fluctuations in one dimension, the groundstate is a (disordered) singlet and
there is a gap, $\Delta$ to the lowest excited states, a triplet. For large
$s$,
\begin{equation} \Delta \propto Je^{-\pi s}.\end{equation}
(In the half-integer $s$ case, an additional topological term must be added to
the Lagrangian, leading to gapless behaviour.)  The staggered spin field, $\vec
\phi$, acting upon the groundstate, produces the triplet of single-particle
states.  However, the uniform magnetisation field, $\vec l$, is a two-magnon
operator, producing or annihilating a pair, or else flipping the polarization
of
a magnon. Hence the low temperature neutron scattering cross-section is
dominated by single magnon production for wave-vector $q$ near $\pi$ and by
two-magnon production for $q$ near $0$.

Unfortunately, relatively little exact information is available about the
$\sigma$-model.  Exact S-matrix results indicate that the spectrum consists
purely of the triplet of magnons, together with multi-particle scattering
states but no boundstates.  The exact matrix element of $\vec l$ between vacuum
and two-magnon states is also known and can be used to predict the $T=0$
neutron-scattering cross-section at small $q$ and energies between $2\Delta$
and $4\Delta$ (where 4-magnon production becomes possible).  It is possible to
add additional terms to the Hamiltonian, reflecting the crystal field
anisotropy:
\begin{equation} {\cal H} \to {\cal H} + a\left(\phi^z\right)^2 +
b\left[\left(\phi^x\right)^2-\left(\phi^y\right)^2\right].\end{equation}
However, no exact results are available on this model.  Results are
available from the large-n and related self-consistent approximations.  In
these
approximations, the fields, $\phi^a$, essentially become free fields with
self-consistently determined masses and repulsive interactions which are
assumed weak [of O(1/n)].

The free, or weakly interacting boson model, sometimes referred to as the
quantum Landau-Ginsburg model \cite{aff1},
is obtained by relaxing the non-linear
constraint, $\vec \phi^2=1$ and writing a simple $\vec \phi^4$ Hamiltonian:
\begin{equation}
{\cal H} = {v\over 2}\vec \Pi^2 + {v\over 2}\left({d\vec \phi \over
dx}\right)^2
+{1\over 2v}\sum_{a=x,y,z}\Delta_a^2\phi_a^2 + \lambda |\vec \phi
|^4.\end{equation}
Here the $\Delta_a$ are
phenomenological mass terms, giving the gaps, which can be given different
values
for each of the three components, $\phi^a$, to model crystal field anisotropy.
A
phenomenological $|\vec \phi|^4$ term is included, when needed for stability,
but
is treated perturbatively.

A third, free (or weakly interacting) fermion model has quite a different
motivation and is not obviously related to either
of the above field theories \cite{tsev}.
The Hamiltonian is written in terms of a triplet of Majorana (ie. Hermitean)
fermion fields, $\vec \psi_{L}$ and $\vec \psi_{R}$.  Here $L$ and $R$ label
left and right-moving components. The Hamiltonian density is:
\begin{equation} {\cal H} = \frac{1}{2} \left [
\vec \psi_L iv{d \over dx}\cdot \vec \psi_L
-\vec \psi_R iv{d\over
dx}\cdot \vec \psi_R + i\sum_a\Delta_a(\psi_{Ra}\psi_{La}-\psi_{La}\psi_{Ra})
+\lambda (\vec \psi_L\times \vec \psi_L)\cdot  (\vec \psi_R\times \vec \psi_R)
\right ] .
\label{Hferm}\end{equation}
  Rather than being based on
the large-s approximation, it appears to be special to the case s=1.  It
presumably becomes exact for the Heisenberg model with an additional isotropic
biquadratic exchange interaction chosen to have the precise value to make the
Haldane gap vanish \cite{aff2}.  (This model is on the phase boundary
separating the Haldane
and spontaneously dimerized phases.)  This model has been shown to be
equivalent
to a $k=3$ Wess-Zumino-Witten (WZW) non-linear $\sigma$-model, or equivalently,
three decoupled critical Ising models.
The critical Ising model in turn, is essentially
equivalent to a massless free Majorana (ie. hermitian) fermion.  Reducing the
biquadratic coupling moves the Ising models away from their critical point,
corresponding to adding mass terms for the fermions.  Four fermion interaction
terms also appear, as written above, but these are generally ignored or treated
perturbatively, in a similar phenomenological spirit to that of the
Landau-Ginsburg boson model.  In this representation, the uniform magnetisation
density, $\vec l$ is again quadratic in the fundamental fields:
\begin{equation} \vec l = \frac{-i}{2} \left (\vec \psi_L\times \vec \psi_L
+\vec \psi_R\times \vec \psi_R \right ).\end{equation}  The particles
created by the fields, $\vec \psi$ are identified as magnons with masses
$\Delta _i$. We see that $\vec l$ is again a two-magnon operator.  It is also
possible to represent the staggered magnetisation, $\vec \phi$ in this
approach,
but it is considerably more complicated.  Near the massless point, this
operator
reduces to the fundamental field of the WZW model, or equivalently to
products of the order and disorder fields, $\vec \sigma$ and $\vec \mu$ of the
three Ising models.  These operators are non-local with respect to the fermion
fields.  The corresponding correlation functions can be expressed in terms of
products of Pain-Lev\'e functions.  They exhibit poles at the fermion masses
together with additional structure at higher energy.  Unlike in the free boson
model, a simple interpretation of the staggered magnetization density as a
single magnon operator doesn't hold.

To summarize, all three models have a triplet of massive magnons.  In the first
two models these are bosons and in the last fermions.  In all models the
uniform magnetisation density, $\vec l$, is a two magnon operator.  In the free
boson or fermion models this is exact; including interactions in any of the
models we expect some contribution from $4$ or more magnons.  In the free boson
model the staggered magnetization, $\vec \phi$, is purely a single magnon
operator.  Including interactions there is an admixture of $3$ and more
magnons.  Recent numerical work indicates that this admixture is very small
\cite{sor}.
This operator has a complicated representation, non-local in the fermionic
magnon fields, in the fermion model.

We will be concerned with adding a magnetic field to the Hamiltonian.  If the
associated magnetization is conserved, then the effect is rather simple.  This
would occur, for example, if the field is in the $3$ direction and rotational
symmetry about the $3$ axis is a good symmetry; ie. $\Delta_1=\Delta_2$.
Consistency demands that, in any of the field theories, we couple the field to
the conserved density, $\vec l$:
\begin{equation} {\cal H} \to {\cal H} -g\mu_B \vec H\cdot \vec
l.\label{field_term}\end{equation} If the associated magnetization is not
conserved, then extra terms could be added \cite{halp}.
We will  use the term in Eq. (\ref{field_term}) for simplicity.  In the case
where the magnetization is conserved, the energy of any state is simply shifted
proportional to its magnetization, $m$, $E \to E-g\mu_BHm$.  In particular, the
magnon energy gaps become:
\begin{eqnarray}
\Delta_3(H) &=& \Delta_3\nonumber \\
\Delta_\pm (H) &=& \Delta_1 \pm g\mu_BH.\end{eqnarray}
Here the last two magnon eigenstates are created by the fields $\phi_1\pm
i\phi_2$.
If the magnetization is not conserved then the field dependence of
the energies becomes model dependent.  Explicit formulas for magnon energies,
$E_a(k,H)$, have been derived in free boson and fermion models \cite{aff}.
The fermion
result seems to agree very well with experiment; the boson one less so.  As the
field increases one of the gaps decreases, eventually reaching zero at a
critical
field of $H \sim O(\Delta /g\mu_B)$.

At low temperatures, we expect the nuclear magnetic relaxation to be dominated
by $q$ near 0 or $q$ near $\pi$ since this is where the lowest energy
processes occur.  It is clear that $1/T_1$ will vanish exponentially as $T\to
0$ since there is a gap.  We wish to investigate whether the dominant
contribution comes from $q\approx 0$ or $q\approx \pi$.  Let us first use the
boson model to consider $q\approx \pi$.  From Eq. (\ref{1/T_1}), we see that:
\begin{eqnarray} 1/T_1 &\propto & \sum_{b,i} A^{bi} A^{ib}
\sum_{n,m}e^{-E_n/T}\left(1+e^{\omega_N/T}\right)<n|\phi^i|m><m|\phi^i|n>
\nonumber \\ && \delta
(E_n-E_m-\omega_N).
\end{eqnarray}

Note that the expression is diagonal in spin operators. This is because we
only consider Hamiltonians that have at least $Z_2 \times Z_2 \times Z_2$
symmetry. Any correlators which contain cross terms of spin operators will
be odd under some $Z_2$ operation and must therefore vanish.

\begin{figure}
\epsfxsize=14 cm
\epsffile{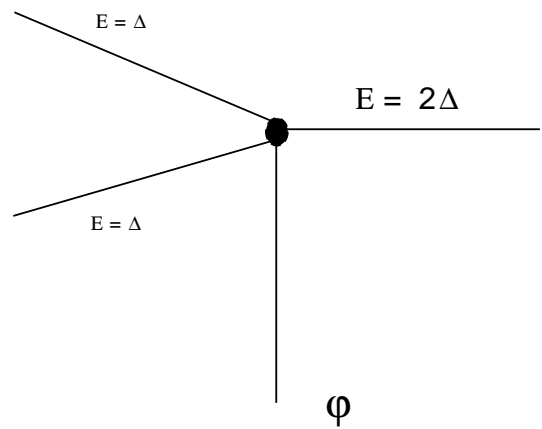}
\caption{First non-vanishing contribution to relaxation due to the staggered
part of the spin}
\label{fig:1}
\end{figure}

  We assume that the magnetic field
is well below $H_c$.  As we have mentioned above, the staggered magnetization
field, $\vec \phi$ is a single magnon operator in the non-interacting boson
model.
Thus it only has matrix elements between states whose energies differ by a
magnon energy, $E_a(k,H)$.  In particular, there is no matrix element with
energy difference $\omega_N$ (which is essentially zero).  Including
interactions
in the boson model (as for instance given by the NL$\sigma$ model) there will
be
a contribution from $q\approx \pi$ at finite $T$. The simplest process is shown
in Fig. \ref{fig:1}.
It involves a $\phi^4$ interaction such as occurs in the quantum
Landau-Ginsburg or NL$\sigma$ models.  The vertical line represents the field
$\phi$.  The incoming line from the right represents a thermally excited magnon
of non-zero momentum, $k$, and energy $2\Delta$.  (For simplicity, we consider
the isotropic model, at zero field.)  The two outgoing lines to the left
represent magnons at rest.
This diagram gives a non-zero matrix element proportional to $
\lambda /\Delta^2$.  Note however, that since the energy of the initial and
final state must be at least $2\Delta$, there will be a Boltzmann suppression
factor of $e^{-2\Delta /T}$ in Eq. (\ref{1/T_1}).  Thus the contribution to
$1/T_1$ will be proportional to:
\begin{equation} 1/T_1 \propto \lambda^2 e^{-2\Delta /T}.\end{equation}
Including anisotropy and a finite field, there will be various contributions of
this type.  It remains true that all these processes are suppressed by
exp$[-2\Delta_{\hbox{min}}(H)/T]$, where $\Delta_{\hbox{min}}(H)$
is the minimum gap.  It is also possible to interpret this
calculation as giving a finite width to the single magnon state at finite T.
However, this cannot change the conclusion that there is a double exponential
suppression factor, contrary to the proposed model in Ref.
(\onlinecite{fuji}).

\begin{figure}
\epsfxsize=14 cm
\epsffile{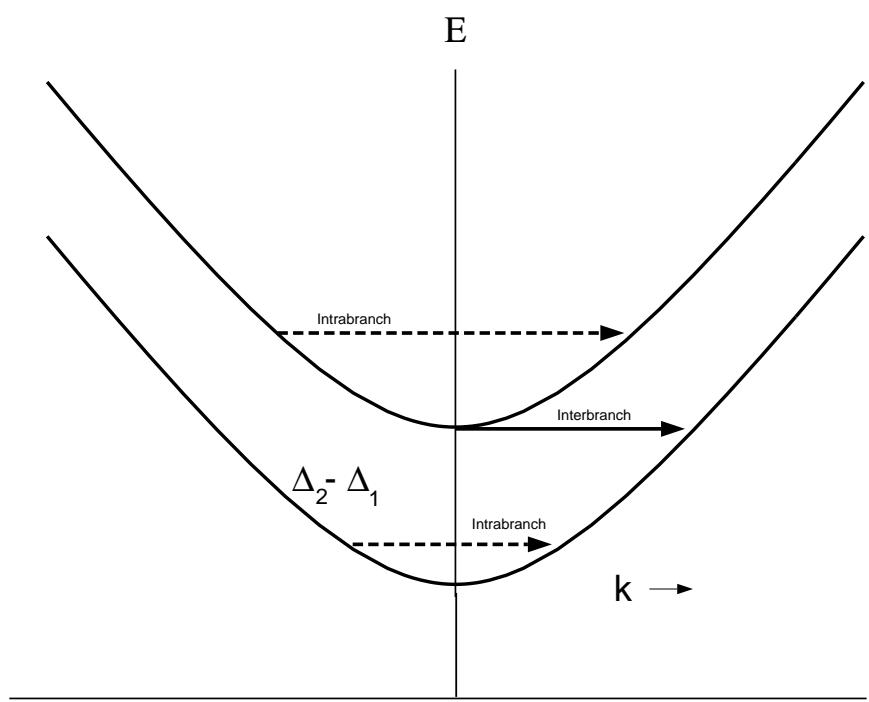}
\caption{Inter- vs Intrabranch transitions}
\label{fig:2}
\end{figure}

Let us now consider contributions to $1/T_1$ from $q\approx 0$.  These are
given by:
\begin{equation} 1/T_1\propto \sum_{b,i}\int_{-\infty}^\infty dt
 A^{bi} A^{ib} e^{-i\omega_Nt}<\{l^i(0,t),l^i(0,0)\}>.\end{equation}
 There is now, in general, a contribution even in the non-interacting boson or
fermion model. This comes from the terms in $l^b$ which annihilate one magnon
and create one magnon.
In the presence of anisotropy, the three magnon branches
are split, so we must distinguish inter-branch and intra-branch transitions
(see Fig. \ref{fig:2}).
Note that both are possible since the wave-vector need not be conserved in the
transition.  However, the Boltzmann factor will be smaller for the inter-branch
transition.  In this case the optimum situation corresponds to a transition
from a lowest branch magnon with a large kinetic energy to a middle branch
magnon with zero kinetic energy.  The Boltzmann factor is then
exp$[-\Delta_{\hbox{middle}}/T]$.  Intra-branch transitions, in order to
approximately conserve energy, will either have essentially zero change in
wave-vector (forward scattering) or else the wave-vector will be essentially
reversed (backscattering).  Clearly the dominant process will be the
intrabranch transition for the lowest branch near the minimum energy $k\approx
0$.  The Boltzmann factor is now exp$[-\Delta_{\hbox{min}}/T]$.  This
contribution will be discussed in detail in the next section.

We now wish to discuss the behaviour near and above the critical field.
Clearly once we are sufficiently close to $H_c$ that the lowest gap is of
$O(T)$, the previous analysis breaks down.  The $q\approx \pi$ part may become
comparable or even larger than the $q\approx 0$ part.  We also need to consider
possible infrared divergences and multi-magnon contributions.  The nature of
the critical point was established in Ref. (\onlinecite{Aff}).  If we
assume exact rotational symmetry around the field direction (which is
approximately true in NENP when the field is in the chain direction) then the
phase
transition is in the two-dimensional $xy$ universality class, corresponding to
Bose condensation of the lowest branch of bosons.  There is quasi-long-range
order for the staggered moment perpendicular to the field direction above
$H_c$, and an associated gapless mode.  Without axial symmetry, the phase
transition is in the 2-dimensional Ising universality class; there is true
long-range-order above $H_c$, with the correlation length and inverse gap only
diverging right at $H_c$.

Rather remarkably, the free fermion
model gives the exact critical behaviour of the uniform part of the spin
operator, with or without axial symmetry.  An argument for this, in the axially
symmetry case, was given in Ref. (\onlinecite{Aff})
[See also Ref.  (\onlinecite{sachdev}).]
This follows from the fact that the many body
wave-function for a very dilute gas of repulsive bosons is just a free fermion
wave-function multiplied by a sign function to correct the statistics.  It is
also fairly easy to see that the free fermion model becomes exact without axial
symmetry. This follows from the well-known equivalence of the Ising model and a
free Majorana fermion near the critical point.  We emphasize that the free
fermion model is certainly not exact far from $H_c$ (although comparison with
experiment indicates that it works well).  Magnon interaction terms such as
that in Eq. (\ref{Hferm}) must be included in general.  However, they become
irrelevant near $H_c$.  This is quite easy to see in the Ising case.  Two of
the three branches of Majorana fermions remain gapped at the critical point and
can be integrated out.  Various interaction terms will be generated in the
effective Hamiltonian for the remaining gapless fermion.  However, there are no
relevant interactions possible for a Majorana fermion since
$\Psi_L\Psi_L\Psi_R\Psi_R$ vanishes by fermi statistics.  Thus we can easily
calculate the $q\approx 0$ part of $1/T_1$ right through the critical point.
This is done in Section IV.  The $q\approx \pi$ part is more difficult since it
requires correlation functions for the 2-dimensional Ising order and disorder
variables away from the critical point on a cylinder of radius $1/T$.  A more
sketchy discussion of this contribution is also given in Sec. IV.  It turns out
that the $q\approx \pi$ part dominates very close to $H_c$.  Well above $H_c$,
in
the axially non-symmetric case, when the gap becomes larger than $T$ again, it
is
simpler to use the weakly interacting boson model, as discussed in Ref.
(\onlinecite{Aff}), to calculate the staggered part.

\section{ NMR Rates Below $H_c$ }

\subsection{ $T_1^{-1}$  With Axial (or $SU(2)$) Symmetry:
Universality of Result, Exact Sigma Model Result.}

We first consider the isotropic case with the uniform magnetic field
in the $z$-direction and the RF field in the $x$-direction.
If the
hyperfine couplings are also isotropic, $A^{ab} \propto \delta^{ab}$, then the
intrabranch process doesn't occur. This follows because the RF field is
perpendicular to the static magnetic field, so $1/T_1$ involves only matrix
elements
of $S^{\pm}$. Since the three magnon branches have definite values of $S^z =
-1,0,1$,
$S^{\pm}$ can only produce interbranch transitions. For more general hyperfine
couplings,
there will also be a contribution from $S^z$. Defining,
\begin{eqnarray}
S^z_q = \frac{1}{N} \sum_x e^{-iqx} S^z(x)
\end{eqnarray}
we see that in this case the leading low temperature
relaxation rate will be determined by $<k+q,-|S^z_q|k,->$ and
$<k+q,-|S^z_{q+2k}|-k,->$.
$|k,->$ denotes the state on the lowest branch with wave-vector k and
$s^z = 1$. Here $q$ must be very
small since $\omega_-(k+q) - \omega_-(\pm k) = \omega_N$.
At low $T$, k must also be quite small since
the Boltzmann factor is $\exp(-\omega_-(k)/T)$.
Since the total $z$ component of spin remains a
good quantum number in the presence of the magnetic field, it follows that
for $k \approx 0$,

\begin{eqnarray}
<k,-|S^z(0)|k,-> = 1
\end{eqnarray}

where we've chosen the following measure for the one particle states,
\begin{eqnarray}
<k|k^{\prime}> = 2\pi \delta(k-k^{\prime}).
\end{eqnarray}

(This is true for arbitrary field,
$H$.) In this case we can give a model independent, exact result
for the low $T$ relaxation rate. We also need the dispersion relation for the
lowest branch
at small k. By analyticity in $k$ this must have the form:
\begin{eqnarray}
\omega_-(k) = \Delta - h + (vk)^2/2\Delta + O(k^4),
\end{eqnarray}
where $\Delta$ is the zero field gap of the $S^z = \pm1$ branches and
we may regard this formula as a definition of $v$; for convenience,
we define
\begin{eqnarray}
h \equiv g\mu_B H.
\end{eqnarray}
In the Lorentz invariant field
theory models, we have:
\begin{eqnarray}
\omega_-(k) = \sqrt{\Delta^2 +(vk)^2} - h,
\end{eqnarray}
consistent with this form. Thus we obtain from (\ref{1/T_1}) the following
expression for intrabranch relaxation along the lowest branch:
\begin{eqnarray}
\left ( 1/T_1 \right )_{\mbox{Intra}} &  \propto & 4\pi|A^{xz}|^2
\int \frac{dk \; dq}{(2\pi)^2}
\delta( \omega_-(k+q) - \omega_-(k) - \omega_N) \nonumber \\
 & \times & e^{-\omega_-(k)/T}
|<k+q,-|S^z(0)|k,->|^2  \label{lbl2} \\
  & \approx & \frac{4|A^{xz}|^2 \Delta e^{-(\Delta-h)/T}}{\pi v^2}
\int_0^{\infty} \frac{dk}{\sqrt{k^2 + 2\omega_N \Delta/v^2}}
e^{-(vk)^2/2\Delta T} \label{Tu1}
\end{eqnarray}
The square root in the denominator of Eq. (\ref{Tu1})
comes from integrating over the delta function.
Note that it is necessary to keep $\omega_N \neq 0$ in this equation since
otherwise there
is a divergence at $k \rightarrow 0$, from the diverging density of states at
the gap.

At low fields, there will be interbranch transitions between the weakly
split branches as well as intrabranch contributions from states with the
``$+$'' quantum number. The Boltzmann suppression will always be that of
the higher branch. The most important of these subdominant contributions
comes from intrabranch transitions along the ``$+$'' branch. The reason
is that low momentum transitions can occur in these processes
causing a square root factor as in Eq. (\ref{Tu1}) to be of order $\omega_N/v$,
ie. close to where the density of states diverges.
The momentum transfer in the
interbranch transitions must be essentially greater than the gap between
the branches. The intrabranch transition rate along the ``$+$'' branch
is identical to Eq. (\ref{Tu1}), albeit with $h \rightarrow -h$ in the
exponential.

In the regime of interest, $\omega_N \ll T \ll \Delta$,  Eq. (\ref{Tu1}) gives:
\begin{eqnarray}
\left ( 1/T_1 \right )_{\mbox{Intra}} \propto |A^{xz}|^2
\frac{4\Delta}{\pi v^2}
[\log(4T/\omega_N) - \gamma](1+e^{-2h/T})
e^{-(\Delta-h)/T}
\label{def}
\end{eqnarray}

where $\gamma=0.577216....$ is Euler's constant.
Essentially this formula, at $h=0$, was given independently by Jolicoeur and
Golinelli
\cite{joli},
based on the large-N approximation to the $NL\sigma$ model. A similar
expression
was also derived by Troyer et. al. for the Heisenberg ladder problem having
an excitation spectrum identical to the isotropic model discussed here
\cite{troy}.
The present derivation shows that
it is an exact result at $T \rightarrow 0$. We note that weak interchain
couplings, $J^{\prime}$,
would also cut off this infrared divergence, replacing $\omega_N$ in the above
expression by a
quantity of order $\sqrt{ J J^{\prime}} $.

Interbranch transitions will correspond to matrix elements
\begin{eqnarray}
|<k+q,\pm|S^x(0)|k,0>|^2 = |<k+q,\pm|S^y(0)|k,0>|^2 \approx \frac{1}{2}
\end{eqnarray}
$k$ and $q$ are assumed small.
The expression for the relaxation rate is identical to Eq. (\ref{lbl2}) with
the appropriate matrix elements substituted and the energy of the higher
branch in the exponential. Energy conservation enforced by the delta
function will now replace $\omega_N$ by $h(1+\frac{h}{2\Delta})$
in Eq. (\ref{Tu1}). One also needs
to multiply the result by an overall factor of two corresponding to
exchanging the labels of the states in the trace. The result is
\begin{eqnarray}
\left ( 1/T_1 \right )_{\mbox{Inter}} \propto (|A^{xx}|^2+|A^{xy}|^2)
\frac{4\Delta}{\pi v^2}
[\log\left(\frac{4T}{h(1+\frac{h}{2\Delta})}\right) -
\gamma](e^{-h/T}+e^{-2h/T})
e^{-(\Delta-h)/T}
\end{eqnarray}

Writing:
\begin{eqnarray}
1/T_1 = \left ( 1/T_1 \right )_{\mbox{Intra}} +
\left ( 1/T_1 \right )_{\mbox{Inter}} \equiv F(h,T) e^{-(\Delta-h)/T}
\end{eqnarray}
following Fujiwara et. al. \cite{fuji}; we see that
$F(h,T)$ decreases rapidly with $h$ until $h\approx T$ at which point
the effects of the upper branches disappear.

At somewhat higher $T$, the $k$ dependence of the matrix elements will become
important,
as will the interbranch transitions.
In the free boson model intrabranch matrix elements are, in fact, independent
of $h$ and identical for forward and backward scattering:
\begin{eqnarray}
<\pm k,-|S^z(0)|k,-> \approx <\pm k,-|l^z(0)|k,-> = 1
\end{eqnarray}
The interbranch matrix elements (corresponding to transitions $-
\leftrightarrow 0$
and $+ \leftrightarrow 0$) will be
\begin{eqnarray}
|<k,\mp|S^{\pm}|k^{\prime},0>|^2 \approx \frac{1}{2}
\left [
\sqrt{\frac{\omega_{k^{\prime}}^{\prime}}{\omega_k}}
+ \sqrt{\frac{\omega_k}{\omega_{k^{\prime}}^{\prime}}} \right ]^2
\end{eqnarray}
where $\omega_k = \sqrt{(vk)^2 + \Delta^2}$, $ \omega_{k^{\prime}}^{\prime}
=  \sqrt{(vk^{\prime})^2 + \Delta^2_0}$ and $\Delta_0$ is the gap to the
$S^z = 0$ branch (see the appendices for more
details). Note that this result is still independent of the sign of $k$.

In the free fermion and $NL\sigma$ models the expression for backward
scattering is somewhat modified:

\begin{eqnarray}
<-k,-|S^z(0)|k,->=G(\theta)
\end{eqnarray}
The fermion model gives
\begin{eqnarray}
G(\theta) =-i\Delta/\omega_k
\end{eqnarray}
while in the $NL\sigma$ model one gets
\begin{eqnarray}
G(\theta)=\frac{\pi^2}{2\theta}
\frac{(\theta-i\pi)}{(\theta-i2\pi)}\tanh(\theta/2)
\end{eqnarray}
where the rapidity variable, $\theta$ is given by
\begin{eqnarray}
\cosh(\theta) = -1 - \frac{2v^2 k^2}{\Delta^2}
\end{eqnarray}
Note that these are still independent of $h$.
For $vk \ll \Delta$, $\theta = i\pi + \frac{2vk}{\Delta}$, and
\begin{eqnarray}
G(\theta) \approx 1 - ( (\frac{2}{\pi})^2 + \frac{1}{3} )(vk/\Delta)^2
\end{eqnarray}
This is effectively due to magnon interactions and will, in turn, contribute a
subleading
$T$ dependence to the prefactor, $F(h,T)$. Estimating the strength of the
magnon interactions
from experimental measurements of $1/T_1$ would be difficult, due to various
other
sources of temperature dependence.

Interbranch transitions in the fermion model are also sensitive to the
sign of the scattering direction:
\begin{eqnarray}
|<\pm k,-|S^{+}|k^{\prime},0>|^2 \approx
\frac{\Delta^2_0}{2\omega_k \omega_{k^{\prime}}^{\prime}} \left [
\sqrt{\frac{\omega_k \mp k}{\omega_{k^{\prime}}^{\prime}+k^{\prime}}} +
\sqrt{\frac{\omega_k \pm k}{\omega_{k^{\prime}}^{\prime}-k^{\prime}}} \right
]^2
\approx |<\pm k,+|S^{-}|k^{\prime},0>|^2
\end{eqnarray}
Notice that those interbranch transitions matrix elements which
contribute to $1/T_1$ are $h$ dependent in
all cases. This comes from the constraint
$\omega_{\pm}(k) = \omega_0(k^{\prime})$.

Actually, all of the above arguments still hold with only $U(1)$ symmetry when
the
field is along the symmetry direction.
In principal, the function $G(\theta)$ will remain independent of field
but depends on how the $SU(2)$
symmetry is broken down to $U(1)$. It is important to emphasize
the universality of the leading term at at low $T$.
The only requirement
is that the system possess $s_z=\pm 1$ degenerate low lying states with
dispersion relation, $\omega \sim \Delta + \frac{v^2 k^2}{2\Delta}$ as $k
\rightarrow 0$.
Once again, the addition of $hS^z$ to the Hamiltonian will not change the
matrix elements of
$l^z$. The new energies will be $\omega \pm h$, and with the low
temperature assumption, one recovers the leading intrabranch contribution
to $1/T_1$.

\subsection{$T_1^{-1}$ With $Z_2$ Symmetry, Subcritical Magnetic Field Results}

Recall that for temperatures much smaller than the lowest gap,
we need only consider the uniform part of the spin, $\vec{l}$.
We are thus led to calculate the transition matrix between two single particle
states
\begin{eqnarray} \vec{l}_{a,b}(k,k^{\prime}) \equiv \int_{-\infty}^{\infty}
\frac{dq}{2\pi} <k,a|\vec{l}_{q}^{i}(t=0) |k^{\prime},b> \end{eqnarray}
where $|k,a>$ and $|k^{\prime},b>$ are magnon states, and $a$ and $b$ denote
the different magnon branches.

The details of the calculation for the free fermion/boson models are given
in the appendices.

       Again, we perform this calculation in the low temperature limit $(\beta
\omega_- \gg 1)$.
In this regime, and with an {\em anisotropic} hyperfine tensor,
we expect the dominant transitions to be
intrabranch unless the branches
cross or come close to doing so (this in fact happens when the field is placed
parallel to
the low branch axis). We start by calculating intrabranch contributions
in both boson and fermion models.
The relevant component of $\vec{l}_{--}$ is along the magnetic field.
Taking this to be the 3 direction and the 1 direction corresponding to
the RF field, Eq. (\ref{lbl2}) becomes
\begin{eqnarray} T_{1_{--}}^{-1} \propto |A^{13}|^2
\frac{4}{(2\pi)}\int_{0}^{\infty}  \frac{\omega_-(k) \; dk}
{\sqrt{k^2+\epsilon(k)^2}}
e^{-\beta \omega_{-}(k) } \nonumber \\
\left ( |l^{3}_{-,-}(k,k)|^{2}+|l^{3}_{-,-}(-k,k)|^{2} \right )
\left ( \frac{\partial \omega^2_-}{\partial k^2}
\right )^{-1}  \label{tz2}
\end{eqnarray}
The function $\epsilon(k)$ is defined by
\begin{eqnarray} \omega_- (k^2+\epsilon(k)^2) = \omega_N + \omega_- (k^2)
\end{eqnarray}
\begin{eqnarray} \epsilon(k)^2 \simeq \omega_N (2\omega_{-}(k^2) + \omega_N)
\left ( \frac{\partial \omega^2_-}{\partial k^2} \right )^{-1}_{k=0}
\label{eps}
\end{eqnarray}
and the three branches have the following dispersion relations

\begin{eqnarray}
\omega_{\pm}^2 = h^2 + v^2k^2 + \frac{(\Delta_1^2 + \Delta_2^2)}{2} \pm
\sqrt{ 4h^2 \left ( v^2k^2 + \frac{(\Delta_1^2 + \Delta_2^2)}{2} \right ) +
\frac{(\Delta_1^2 - \Delta_2^2)^2}{4} } \nonumber
\end{eqnarray}
\begin{eqnarray}
\omega_3^2 = v^2k^2 + \Delta_3^2 \mbox{\hspace*{2.0in} Bosons}
\label{enerb} \end{eqnarray}

\begin{eqnarray}
\omega_{\pm}^2 = h^2 + v^2k^2 + \frac{(\Delta_1^2 + \Delta_2^2)}{2} \pm
\sqrt{ 4h^2 \left ( v^2k^2 + \frac{(\Delta_1 + \Delta_2)^2}{4} \right ) +
\frac{(\Delta_1^2 - \Delta_2^2)^2}{4} } \nonumber
\end{eqnarray}
\begin{eqnarray}
\omega_3^2 = v^2k^2 + \Delta_3^2 \mbox{\hspace*{2.0in} Fermions}
\label{enerf}
\end{eqnarray}

When $\beta \omega_- \gg 1$ the integral in Eq. (\ref{tz2}) is strongly
peaked at $k=0$. We can then make the following approximation (the validity
of this will be discussed in a larger context later in this section):
\begin{eqnarray} T_{1_{--}}^{-1}
\propto 8 |A^{13}|^2 C e^{-\beta \omega_{-}(0)} |l^{3}_{-,-}(0,0)|^{2}
\frac{\omega_-(0)}{2\pi} \left ( \frac{\partial \omega^2_-}{\partial k^2}
 (0)   \right )^{-1}
\label{eq1}
\end{eqnarray}

\begin{eqnarray} C = \int_{0}^{\infty} dk \frac{e^{\beta (\omega_{-}(0)-
\omega_{-}(k) )}}
{\sqrt{k^2+\epsilon(k)^2}}
\end{eqnarray}
\[ \approx \int_{0}^{\infty} dk
\frac{e^{-\beta \left ( \frac{\partial \omega^2_-}{\partial k^2} \right )
\frac{k^2}{2\omega_{-}(0)} }}{\sqrt{k^2+\epsilon(k)^2}} \]
Using Eq. (\ref{eps}) and making the change of variables
$\beta \left ( \frac{\partial \omega^2_-}{\partial k^2} \right )
\frac{k^2}{2\omega_{-}(0)} \rightarrow k$, we get
\begin{eqnarray}
C = \int_{0}^{\infty} dk \frac{e^{-k^2}}{\sqrt{ k^2 + \omega_N (2\omega_{-}(0)
+ \omega_N) \beta/2
\omega_{-}(0) }} \nonumber \\
 = e^{\omega_N (2\omega_{-}(0) + \omega_N) \beta/4\omega_{-}(0)}
K_0 \left ( \omega_N (2\omega_{-}(0) + \omega_N) \beta/4\omega_{-}(0) \right )
\end{eqnarray}

where $K_0$ is the zero order modified Bessel Function.
For $\omega_{-}(0) \gg \omega_N$, which is really the case we are considering,
this reduces to
\begin{eqnarray} C = -\log(\beta \omega_N/4) - \gamma.
\end{eqnarray}
This is identical to the factor arising in Eq. (\ref{def}).

For the boson model (see Appendix \ref{appA}),
\begin{eqnarray}
|l^{3}_{-,-}(0,0)|^{2} = \frac{h^2}{\omega_-^2} \left ( \frac{(\Delta_1^2 +
\Delta_2^2) - \sqrt{2h^2 (\Delta_1^2 + \Delta_2^2) + (\Delta_1^2 -
\Delta_2^2)^2 /4} }
{\sqrt{2h^2 (\Delta_1^2 + \Delta_2^2) + (\Delta_1^2 - \Delta_2^2)^2 /4} }
\right )^2 \end{eqnarray}
so $F(h,T)$ goes to zero as a quadratic
power of the field.
\begin{eqnarray} |l^{3}_{-,-}(0,0)|^{2} \stackrel{h \rightarrow 0}{\rightarrow}
\frac{h^2}{\Delta_2^2} \left ( \frac{\Delta_1^2 + 3\Delta_2^2}{ \Delta_1^2 -
\Delta_2^2} \right )^2
\end{eqnarray}

Where we assumed without loss of generality, $\Delta_2 < \Delta_1$.
For the fermion model,
\begin{eqnarray} |l^{3}_{-,-}(0,0)|^{2} =
\frac{4h^2}{(\Delta_1-\Delta_2)^2 + 4h^2}
\end{eqnarray}
This also vanishes with $h^2$.
\begin{eqnarray} |l^{3}_{-,-}(0,0)|^{2} \stackrel{h \rightarrow 0}{\rightarrow}
\frac{4h^2 \Delta_2^2}{(\Delta_1 - \Delta_2)^2}
\end{eqnarray}

We can also extend the above approximations to intrabranch transitions along
the ``$+$'' branch and interbranch processes. The former amounts to letting
$- \rightarrow +$:
\begin{eqnarray} T_{1_{++}}^{-1}
\propto  8 |A^{13}|^2 e^{-\beta \omega_{+}(0)} |l^{3}_{+,+}(0,0)|^{2}
\frac{\omega_+(0)}{2\pi} \left ( \frac{\partial \omega^2_+}{\partial k^2}
(0)   \right )^{-1} (-\log(\beta \omega_N/4) - \gamma)
\label{eq2}
\end{eqnarray}

The analogous expression for the interbranch rate is slightly more involved.
Assuming that the branch labeled by $r$ lies higher than that labeled by
$s$, we can use the same approximations to arrive at
\begin{eqnarray}
\begin{array}{lll} T_{1_{rs}}^{-1} &
\propto  & 8 e^{-\beta \omega_r(0)} \sum_i |A^{1i}|^2 \left [
|l^{i}_{r,s}(0,Q)|^{2} + |l^{i}_{r,s}(0,-Q)|^{2} \right ]
\frac{\omega_r(0)}{2\pi} \left ( \frac{\partial \omega^2_s}
{\partial k^2}(Q)   \right )^{-1/2} \left ( \frac{\partial \omega^2_r}
{\partial k^2}(0)   \right )^{-1/2}     \\
 & & e^{\frac{Q^2 \left (
\frac{\partial \omega^2_s}{\partial k^2}(Q)   \right )}{4 \omega_r(0) T}}
K_0( \frac{Q^2 \left (
\frac{\partial \omega^2_s}{\partial k^2}(Q)   \right )}{4 \omega_r(0) T} )
\label{eq3}
\end{array}
\end{eqnarray}
$Q$ is defined by $\omega_r(0) = \omega_s(Q)$.
Notice that since $S^3$ is no longer a conserved operator, hence interbranch
transitions $+ \leftrightarrow -$ are allowed.

Once more we define
\begin{eqnarray}
T_1^{-1} =  \sum_{a b} T_{1_{ab}}^{-1}
\equiv  F(h,T) e^{-\beta \omega_-(0)}
\end{eqnarray}
where one is careful not to double count in summing over $a$ and $b$.

There are regimes where it is no longer sufficient to calculate
only one-particle transitions.
When the temperature becomes comparable to the gap,
multiparticle transitions become important.
To include multiparticle transitions
we simply replace the Boltzmann factor by appropriate occupation factors:
$f_b (1+f_b )= \mbox{cosech}^2(\frac{\beta \omega}{2}) /4$ for bosons,
and $f_f (1-f_f )= \mbox{sech}^2(\frac{\beta \omega}{2}) /4$ for fermions.
\cite {mah}
At these temperatures it is also
necessary to include the $k$-dependence of the integrand past the peak at the
origin. We expect that at
temperatures $T \approx \frac{\Delta}{3}$ and fields $h \approx \frac{2\Delta}
{3}$ the numerically integrated results would differ from Eqns.
(\ref{eq1}--\ref{eq3}) by about 10 percent.
Since we have not yet obtained exact expressions for the contribution
of the staggered part of the spin to the relaxation rate and since, in the
regime in question, single magnon contributions are expected to compete
with those from the uniform part of the spin (this will be demonstrated in
the next section), we will not bother to present numerically integrated
results in this paper.

\begin{figure}
\epsfxsize=14 cm
\epsffile{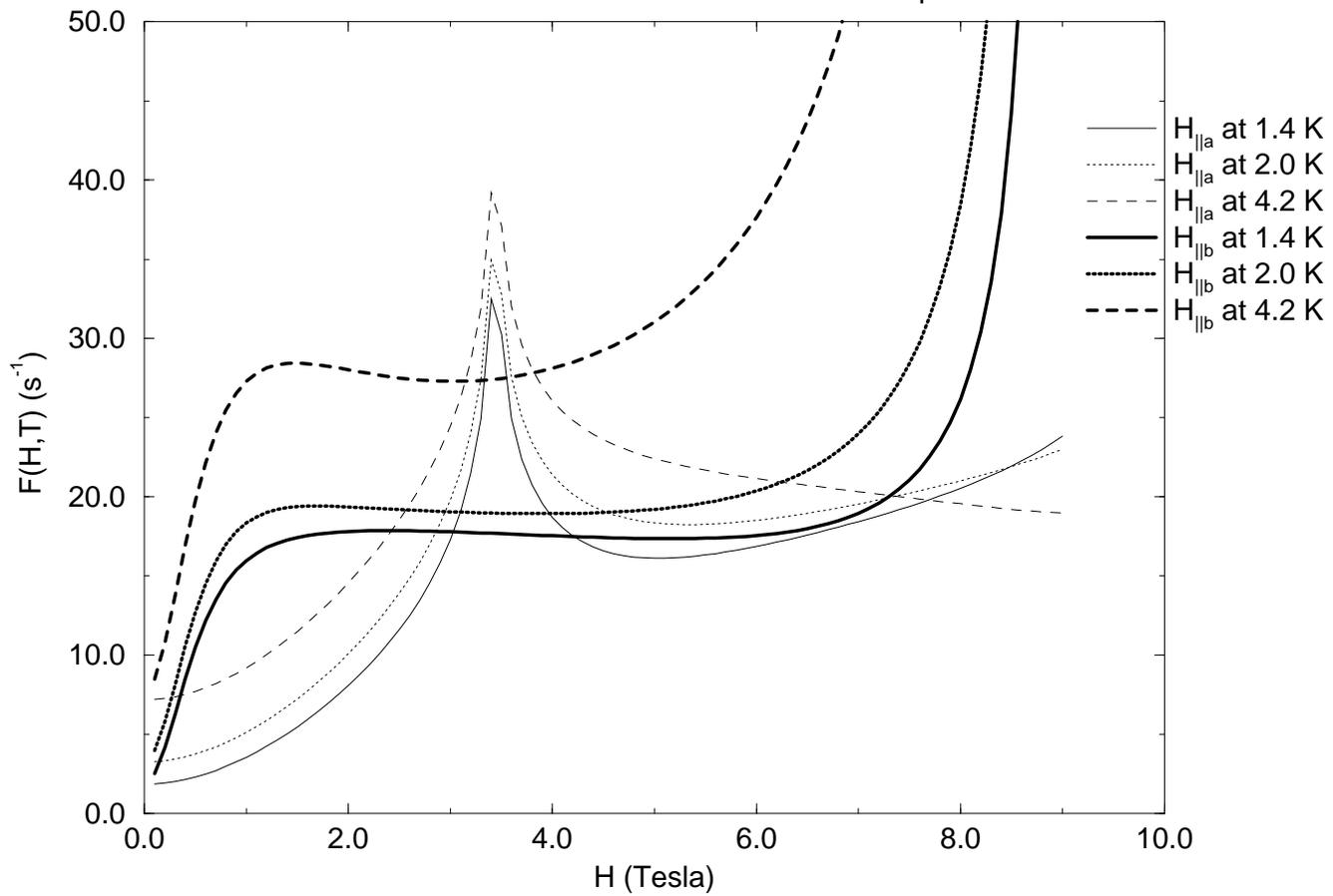}
\caption{F(h,T) for bosons: field along crystal $b$ and $c$ directions}
\label{fig:3}
\end{figure}

\begin{figure}
\epsfxsize=14 cm
\epsffile{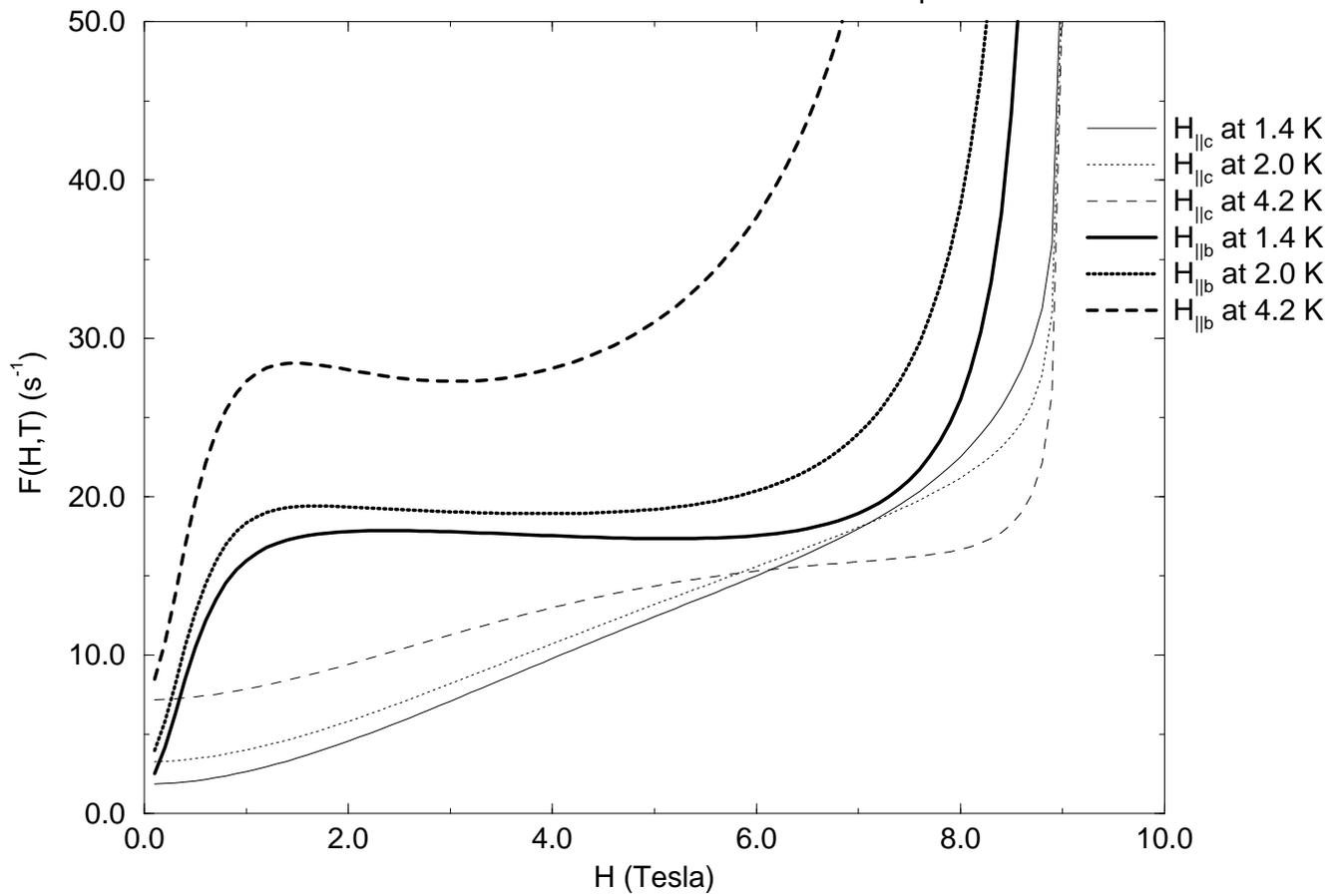}
\caption{F(h,T) for bosons: field along crystal $b$ and $a$ directions}
\label{fig:4}
\end{figure}

\begin{figure}
\epsfxsize=14 cm
\epsffile{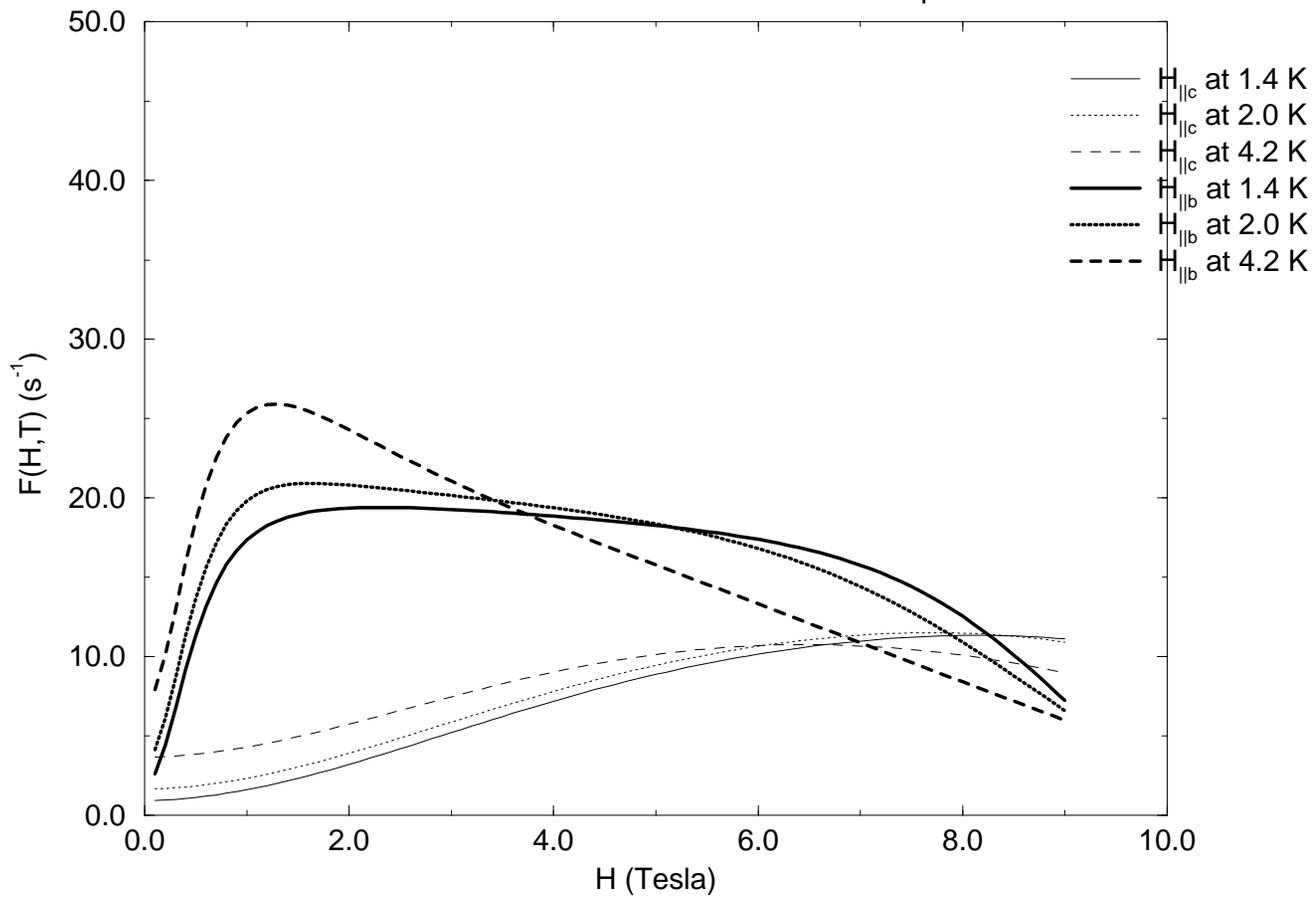}
\caption{F(h,T) for fermions: field along crystal $b$ and $a$ directions}
\label{fig:6}
\end{figure}

In Figures \ref{fig:3}--\ref{fig:6}
we use Eqns. (\ref{eq1},\ref{eq2} and \ref{eq3}) to
plot $F(h,T)$ for bosons and fermions and for fields
along the crystal $a,b$ and $c$ directions in NENP.
We use $\Delta_b = 2.52$mev, $\Delta_a = 1.17$mev
and $\Delta_c = 1.34$mev, and assume a uniform hyperfine coupling
to all spin components on a given site.
We do this for varying temperatures and try to
consistently account for contributions from relevant transitions including
interbranch and multiparticle effects.
Within approximations used, multiparticle effects amount to multiplying
the final expressions by $(1 \pm e^{-\beta \omega_s})^{-2}$.

$F(h,T)$ is shown for fields up to 9 Tesla even though
the $(\beta \omega_- \gg 1)$ approximation is no longer valid at such fields.
This is done
to contrast the predictions of the boson and fermion models. It's easy to see
that the
boson result for $F(h,T)$ diverges at the critical field, while no such
catastrophe is present in
the fermion result. This divergence is logarithmic and infrared. It will
persist even after
account is made for the staggered part of the correlation function.
Multiparticle scattering
will in fact worsen the effect.

In NENP, when the field is along the $b$ direction, we expect relevant
interbranch transitions only
for small field. In this regime, one must also be careful to include
intrabranch
transitions in the second lowest branch. All these processes are of the
same order. Even though the intrabranch rates vanish at low fields, the
interbranch contributions are suppressed by the absence of low momentum
transitions (ie. $Q$ for the interbranch transitions is $O(\sqrt{\Delta^2_1-
\Delta_2^2})$ as opposed to $O(\omega_N)$.)
For this case, only $l^3$ need be calculated.

When the field is along the $c$ direction (corresponding to the middle gap), we
restrict
ourselves to calculating intrabranch transitions along the lower branch and
interbranch
 ones between the lower and $c$ branch. There are no
intrabranch processes along the $c$ axis. Calculating the interbranch
transitions
amounts to calculating
$ |l^{1}_{a,c}|^2 $ and
$|l^{2}_{a,c}|^2 $.

When the field is along the $a$ direction, the calculation proceeds as above.
The
crossing of the branches provides for an interesting effect.
At the subcritical field where branch crossing occurs there is an
infrared divergence in the density of states (ie. $\epsilon(k) \propto k$ for
small $k$).
We expect this to be suppressed by the finite
width of the applied uniform field or by a cutoff corresponding to
interchain coupling but a peak should be seen in $T_1^{-1}$ at this point.
This peak can also be used to locate the true $ac$-axes for the chain
(however, effects discussed in the appendices may broaden this peak
considerably).

As a brief aside, we would like to note that there are some curious differences
in the
behaviour of the boson matrix elements vs.
that of the fermions. We have already seen that some transitions are
fundamentally different
in the two models even at low or zero fields. These may suggest experiments to
further
explore the nature of the low energy excitations in anisotropic Haldane gap
materials.
Even in the simple case of intrabranch backscattering the two theories differ
(the fermion theory predicts a smaller, momentum dependent result).
As another example, consider
\begin{eqnarray}
<0,-|S^i_{q=0}|0,3>
\end{eqnarray}
ie. zero momentum transitions between the low branch and that parallel to the
field (such a
matrix element may be relevant in calculating ESR transition rates).
The fermion model predicts
a nearly field and gap independent result for such a rate
in the limit $h \rightarrow 0$. The boson model
prediction, however, is strongly dependent on the ratio of the
gaps and therefore on the
orientation of the field.
To some degree, the reason that the theories feature such differences {\em even
in the isotropic case}, is that the generator of rotational symmetry is
fundamentally different in the two theories: the fermion magnetization does not
couple
left and right moving fermions, while the boson operator does. In particular,
this
means that intrabranch backscattering should vanish with the fermion mass,
while no
such miracle occurs with the bosons.

There are some puzzling discrepancies between our result for $F(h,T)$
and that of Ref. (\onlinecite{fuji}). At about 5 Tesla,
the experimental data gives $F_b < F_c$ while our theoretical
calculation gives the reverse at low temperatures.
Moreover, the `flatter' of the experimental
curves is the one corresponding to the field perpendicular to the chain
axis (about which there is approximate $U(1)$ symmetry).
The behaviour of the theories is quite easy to
understand from the universal results valid in the axially symmetric case,
discussed
in Section (III.A). $F_b$ is roughly independent of field with axial symmetry
since
$l^3_{--}$ is nearly $h$ independent (in fact, $F_b$ exhibits a logarithmic
divergence
as $h \rightarrow 0$). On the other hand, $F_c$ vanishes quadratically as
$h \rightarrow 0$. Including the small breaking of the axial symmetry
corresponding
to $\Delta_c - \Delta_a = 2^{\circ} K$, $F_b$ is essentially constant down to
low
fields of order $\Delta_c - \Delta_a \approx 1$T, before rapidly decreasing as
seen in
in the figures. Given the original Hamiltonian, Eq. (\ref{eq:h}), we expect
that our calculations are at least qualitatively correct.

Finally, we would like to mention some recent NMR data collected on the
1-D $S=1$ spin chain $AgVP_2S_6$ by Takigawa et. al. \cite{taki}. This material
is highly one dimensional with a large gap ($\Delta \sim 320^{\circ} K$)
and very nearly isotropic ($\delta \sim 4^{\circ} K$). These
characteristics make it ideal for analysis using our results. There are,
however, some questions about the properties of the material which would
have to be analyzed before an understanding of the NMR results is possible
within the framework proposed here. The gap deduced from studies on the
Vanadium atom ($\Delta \sim 410^{\circ} K$)
conflicts significantly with those performed on the Phosphorous
sites and with neutron scattering data. In addition, the material has
very low symmetry (corresponding to the space group $P2/a$) and very
little is known about the possible small $E$ and $D$ terms in the Hamiltonian
and their corresponding symmetry. There is fair qualitative agreement
between the $ ^{31}P$ NMR data and our theory, and it is possible to explain
some of the discrepancies using a temperature dependent anisotropic gap
structure, but we feel that not enough is yet understood about gross
features of the material to justify such speculation at this time.

\section{Close to the Critical Field}

In this final section we would like to demonstrate that the staggered
correlator
contribution becomes crucial as $h \rightarrow h_c$ from below.
First, we would like to describe the transition rate due to the uniform part of
the spin-spin
correlation close to the critical field, $h_c - h \ll k_b T$. In this limit,
the long
wavelength modes dominate the physics, and only intrabranch processes need be
considered.
Since the fermion model becomes exact in this limit, we will rely on its
predictions.
In the $Z_2$ case, the dispersion relation becomes
\begin{eqnarray}
\omega(k,h) = \sqrt{ v_e^2 k^2 + \Delta_e^2}
\end{eqnarray}
where
$v_e^2 = v^2 \frac{(\Delta_1 - \Delta_2)^2}{(\Delta_1 + \Delta_2)^2}$ and the
gap is
$\Delta_e^2 = (h-h_c )^2 \Delta_1 \Delta_2 / (\Delta_1+\Delta_2)^2$.

The $U(1)$ case is different: the dispersion relation in the long wavelength
limit is
\begin{eqnarray}
\omega(k,h) = | (h_c-h) + \frac{v^2 k^2}{2h_c} |
\end{eqnarray}
In the fermion model, the coupling of the magnetic field to a {\em conserved}
charge serves as a
chemical potential for the magnon states. At criticality the dispersion is
quadratic
as opposed to linear as in the $Z_2$ case. Above $h_c$ the chemical potential
drops
below zero indicating that the ground state is now magnetic and filled with
quasiparticles.

In calculating the uniform spin contribution we must now include the
appropriate
occupation numbers in Eq. (\ref{tz2}):
\begin{eqnarray}
\left ( T_1^{-1} \right )_{\mbox{Unif}} \propto \frac{4|A^{13}|^2}{2\pi}
 \int_0^{\infty} dk \frac{ \omega f_f(\omega) (1-f_f(\omega))}
 {\sqrt{k^2 + \epsilon(k)^2 } }
 \left ( \frac{\partial \omega^2}{\partial k^2}  \right )^{-1}
\left ( |l^{3}_{-,-}(k,k)|^{2} + |l^3_{-,-}(k,-k)|^2 \right )
\end{eqnarray}

   In the $U(1)$ case we get a similar equation to (\ref{Tu1}) but with
appropriate
occupation factors. The behaviour is
logarithmic with temperature at $T \gg \omega_N$ even at the critical point.

In the $Z_2$ case we get
\begin{eqnarray}
\left ( T_1^{-1} \right )_{\mbox{Unif}}
\approx \frac{2|A^{13}|^2}{\pi v^2} \frac{\Delta_1 \Delta_2}{(\Delta_1 -
\Delta_2)^2}
\int_0^{\infty}dk \frac{(\omega+\omega_N)
\; \mbox{cosech}^2(\frac{\beta \omega}{2}) }
{\sqrt{ (\omega+\omega_N)^2 - \Delta_e^2} }
\end{eqnarray}

At criticality, we set $\omega \propto k$ and $\Delta_e = 0$.
We may simply rescale the integration variable to obtain
\begin{eqnarray}
\left ( 1/T_1 \right )_{\mbox{Unif}} \propto T
\end{eqnarray}
in the limit
that $\omega_N \ll T$.
 This is expected from the Ising model where the uniform part of the spin
corresponds to the Ising energy density operator, $\epsilon$, of scaling
dimension 1.
In terms of Majorana fermions this
operator is $\psi_L \psi_R = \epsilon$.

We can also say something about the behaviour of the staggered part of the
correlation
function at criticality. In both $U(1)$ and Ising cases we know the critical
behaviours of the staggered spin correlators.
In both cases, the magnetic field acts as the temperature
(and the temperature acts as the Euclidean time interval in an analogous 2D
classical system).
On the infinite Euclidean plane
these correlators are \cite{Aff}:
\begin{eqnarray}
<\phi^{\dagger}(z)\phi(0)> \sim 1/|z|^{\frac{1}{2}}
\end{eqnarray}
\begin{eqnarray}
<\sigma(z) \sigma(0)> = 1/|z|^{\frac{1}{4}}
\end{eqnarray}
 The field $\phi = \phi^x + i\phi^y$ is the charged $U(1)$ field of the boson
model;
$\sigma$ is the disorder field of the Ising model (highly non-local in
fermionic
language) -- it is the only local non-trivial primary
operator aside from the energy density
$\epsilon$.
To get the finite temperature result, we simply make the conformal
transformation from
the plane into the finite strip \cite{car}, $z_p = e^{2\pi iz_s/\beta}$ to get
\begin{eqnarray}
<\sigma(z) \sigma(0)> = \frac{(\pi/\beta)^{\frac{1}{4}}}{|\sin(\pi z
/ \beta)|^{\frac{1}{4}} } \nonumber \\
<\phi^{\dagger}(z)\phi(0)> = \frac{(\pi/\beta)^{\frac{1}{2}}}{|\sin(\pi z
/ \beta)|^{\frac{1}{2}} }
\end{eqnarray}

To get the contribution to $T_1^{-1}$ we integrate over $\int dt e^{i\omega_N
t}$ and set
$z=it+\epsilon$.
\begin{eqnarray}
\left ( T_1^{-1} \right )_{\mbox{Stag}} \propto \int_{-\infty}^{\infty} dt
\frac{e^{i\omega_N t} }{\beta^{\frac{1}{x}}
|\sin(\pi z/ \beta)|^{\frac{1}{x}} }
\end{eqnarray}
$x$ is either 4 or 2.

Since this is analytic as $\omega_N \rightarrow 0$, we get that in the
experimentally important
limit $T \gg \omega_N$
\begin{eqnarray}
\left ( T_1^{-1} \right )_{\mbox{Stag}}
\propto T^{-3/4} + O(\omega_N) \mbox{\hspace*{.3in} Ising case} \nonumber \\
\left ( T_1^{-1} \right )_{\mbox{Stag}}
\propto T^{-1/2} + O(\omega_N) \mbox{\hspace*{.3in} U(1) case}
\end{eqnarray}

In the low $T$ limit,
this will be significantly stronger than the contribution from the uniform
part of the correlator in both theories.
In fact, as long as we are sufficiently close to the
critical regime the above results will only be suppressed by
factors of order $O(\Delta/T)$, where $\Delta$ is the gap and $\Delta \ll T$.
Thus in this regime,
even slightly away from criticality, we see that the most important
contribution
to $1/T_1$ comes from $q=\pi$, ie. the staggered part of the spin correlation
function.

Farther still from criticality, the analysis breaks down but we expect the
staggered spin contribution to influence $1/T_1$ through the region
$T = O(\Delta(h))$.

\vspace*{.2in}

\centerline {\bf Acknowledgements}

We thank WJL Buyers and E. S\o renson for helpful discussions.
J.S. would like to thank M. Walker and his research group at the University
of Toronto where some of this work was done. This research was supported in
part
by NSERC.

\appendix

\section{Diagonalizing the Boson/Fermions Models}
\label{appA}

To do the necessary calculations we need to have a basis of eigenstates
for the Hamiltonian and know the expansion of the field operators
 in terms of creation/annihilation operators for these
states. Obtaining this is tedious (especially when the field does
not lie in a direction of symmetry, for then all the branches mix)
but the idea is to find the right Bogoliubov transformation which
diagonalizes $H$. We do this first for the Boson model.

As discussed in [\onlinecite{bla}], we seek to find the
eigenvectors of the (non-hermitian) matrix $\eta M$, where
\begin{eqnarray} H_{k} =  \vec{a}^{\dagger}_{k} {\bf A} \vec{a}_{k} +
 \vec{a}^{\dagger}_{-k} {\bf A}^{\ast} \vec{a}_{-k} +
  \vec{a}^{\dagger}_{k} {\bf B} \vec{a}^{\dagger}_{-k} +
   \vec{a}_{k} {\bf B}^{\ast} \vec{a}_{-k} \end{eqnarray}
   \begin{eqnarray} \eta M = \left ( \begin{array}{cc}
   {\bf A} & {\bf B} \\
   -{\bf B}^{\ast} & -{\bf A}^{\ast} \end{array}
    \right )   \end{eqnarray}

We will now find the eigenvectors in the case where the field
lies in a direction of symmetry. In this case, only the excitations
transverse to the direction of the applied field mix and we need
only solve a $(4 \times 4)$ set of equations.
Considering the two mixed components of $\vec{\phi}$, we
write
\begin{eqnarray}  \phi^{i}(k,t=0) = \frac{1}{\sqrt{2\omega_{0}}} [a^{i
\dagger}_
{-k}
 + a^{i}_{k}] \end{eqnarray}
  \begin{eqnarray}  \pi^{i}(k,t=0) = i\sqrt{\frac{\omega_{0}}{2}} [a^{i
\dagger}_
  {-k}
    - a^{i}_{k}] \end{eqnarray}
\begin{eqnarray}
[a_k^i, a_{k^\prime}^{j \dagger}] = 2\pi \delta_{ij} \delta(k-k^\prime)
\end{eqnarray}
$\omega_0$ is arbitrary.

Plugging this into the Hamiltonian, we then have that
\begin{eqnarray}
{\bf A} = \frac{\omega_{0}}{4} {\bf I} + \frac{ {\bf K} }{4\omega_{0}} -
\frac{1}{2} h {\bf \sigma}_{2}
\end{eqnarray}
\begin{eqnarray}
{\bf B} = -\frac{\omega_{0}}{4} {\bf I} + \frac{ {\bf K} }{4\omega_{0}}
\end{eqnarray}
 where ${\bf \sigma}_{2}$ is the usual Pauli matrix, and
 \begin{eqnarray} {\bf K} = \mbox{diag} (\Delta_{1}^{2} + v^{2}k^{2},
\Delta_{2}^{2}
 + v^{2}k^{2}) \end{eqnarray}

Summarizing the above, we need to solve:
\begin{eqnarray}   0 = \left ( \begin{array}{cc}
\frac{\omega_{0}}{4} {\bf I} + \frac{ {\bf K} }{4\omega_{0}} -
\frac{1}{2} h {\bf \sigma}_{2} - \frac{\omega}{2} {\bf I} &
	 -\frac{\omega_{0}}{4} {\bf I} + \frac{ {\bf K} }{4\omega_{0}} \\
	 \frac{\omega_{0}}{4} {\bf I} - \frac{ {\bf K} }{4\omega_{0}} &
	 -\frac{\omega_{0}}{4} {\bf I} - \frac{ {\bf K} }{4\omega_{0}} -
	 \frac{1}{2} h {\bf \sigma}_{2} - \frac{\omega}{2} {\bf I}
	 \end{array} \right ) \end{eqnarray}

which can be manipulated to give

\begin{eqnarray}   0 = \left ( \begin{array}{cc}
  (\omega_{0} - \omega) {\bf I} - h {\bf \sigma}_{2} &
     -(\omega_{0} + \omega) {\bf I} - h {\bf \sigma}_{2} \\
	  \frac{ {\bf K} }{\omega_{0}} - \omega {\bf I} - h {\bf \sigma}_{2} &
		 \frac{ {\bf K} }{\omega_{0}} + \omega {\bf I} + h {\bf \sigma}_{2}
				\end{array} \right ) \end{eqnarray}
The eigenvalues of $\eta M$ are already known, as they are the
solutions to the classical equations of motion and come in pairs $\pm
\omega_{\pm}$ (see eqn. \ref{enerb}).
Furthermore, we need only work to find one eigenvector of each
pair because if $\left ( \begin{array}{c} X \\
Y \end{array} \right )$ is a right-eigenvector
of $\eta M$ with eigenvalue $\omega$, then
$\left ( \begin{array}{c} Y^{\ast} \\
X^{\ast} \end{array} \right )$ is a right-eigenvector of $\eta M$ with
eigenvalue $-\omega$ ($X$ and $Y$ are themselves two-component vectors).

The bottom rows give the following set of equations

\begin{eqnarray}  0 = (\Delta_{1}^{2} + v^{2}k^{2}) \chi_{\pm,1}
- \omega_{0} \omega_{\pm} \xi_{\pm,1} +ih\omega_{0} \xi_{\pm,2} \end{eqnarray}
\begin{eqnarray}  0 = (\Delta_{2}^{2} + v^{2}k^{2}) \chi_{\pm,2}
- \omega_{0} \omega_{\pm} \xi_{\pm,2} -ih\omega_{0} \xi_{\pm,1} \end{eqnarray}
where it turns out to be convenient to work with $\chi \equiv X +
Y$ and $\xi \equiv X - Y$. The top rows can be
manipulated to give
\begin{eqnarray} 0 = (h^{2}-\omega_{\pm}^{2}) \chi_{\pm,2} - ih \omega_{0}
\xi_{\pm,1} +
\omega_{0} \omega_{\pm} \xi_{\pm,2} \end{eqnarray}
\begin{eqnarray} 0 = -(h^{2}-\omega_{\pm}^{2}) \chi_{\pm,1} - ih \omega_{0}
\xi_{\pm,2} -
  \omega_{0} \omega_{\pm} \xi_{\pm,1} \end{eqnarray}

These can then be worked to give
\begin{eqnarray} (\Delta_{1}^{2} + v^{2}k^{2} + \omega_{\pm}^{2} - h^{2})
\chi_{\pm,1} =
2\omega_{0} \omega_{\pm} \xi_{\pm,1} \end{eqnarray}
\begin{eqnarray} (\Delta{_2}^{2} + v^{2}k^{2} + \omega_{\pm}^{2} - h^{2})
\chi_{\pm,2} =
2\omega_{0} \omega_{\pm} \xi_{\pm,2} \end{eqnarray}
\begin{eqnarray} (h^{2} + \Delta_{1}^{2} + v^{2}k^{2} - \omega_{\pm}^{2})
\chi_{\pm,1} =
-2ih\omega_{0} \xi_{\pm,2} \end{eqnarray}
\begin{eqnarray} (h^{2} + \Delta_{2}^{2} + v^{2}k^{2} - \omega_{\pm}^{2})
\chi_{\pm,2} =
2ih\omega_{0} \xi_{\pm,1} \end{eqnarray}
Note that if we fix the phase of $X_{1}$ to be real then $Y_{1}$ must
also be real as $X_{2}$ and $Y_{2}$ must be pure imaginary. The normalization
condition now allows us to solve for the eigenvectors which form the columns
of the transformation matrix between the old and diagonal basis of
creation/annihilation operators. This normalization condition is slightly
unusual since $\eta M$ is not hermitean: $X^{\dagger}X - Y^{\dagger}Y = 1$. In
terms of $\chi$ and $\xi$ this is
\begin{eqnarray} \chi_{\pm,1} \xi_{\pm,1} - \chi_{\pm,2} \xi_{\pm,2} = 1
\end{eqnarray}
The solution is

\begin{eqnarray} \chi_{\pm,1} = \left (  \frac{ \omega_{0} \omega_{\pm}
(h^{2} + \Delta_{2}^{2} + v^{2}k^{2} - \omega_{\pm}^{2}) }
{(\Delta_{1}^{2} + v^{2}k^{2} + \omega_{\pm}^{2} - h^{2})
(h^{2} + \frac{\Delta_{1}^{2} + \Delta_{2}^{2}}{2} + v^{2}k^{2}
- \omega_{\pm}^{2}) } \right )^{1/2} \end{eqnarray}
\begin{eqnarray} \xi_{\pm,1} = \frac{\Delta_{1}^{2} + v^{2}k^{2} +
\omega_{\pm}^{2} - h^{2}}
{2\omega_{0} \omega_{\pm}} \chi_{\pm,1} \end{eqnarray}
\begin{eqnarray} \chi_{\pm,2} = \frac{2ih \omega_{0}}
{h^{2} + \Delta_{2}^{2} + v^{2}k^{2} - \omega_{\pm}^{2}} \xi_{\pm,1}
\end{eqnarray}
\begin{eqnarray} \xi_{\pm,2} = \frac{\Delta_{2}^{2} + v^{2}k^{2} +
\omega_{\pm}^{2} - h^{2}}
{2\omega_{0} \omega_{\pm}} \chi_{\pm,2} \end{eqnarray}

We can define the matrices $\chi$ and $\xi$ with the rows labeled by the
eigenvalues
$(+,-)$ and the columns labeled by the original masses $(1,2)$.
One easily verifies that in the limit $h \rightarrow 0$ these
become
\begin{eqnarray} \chi = \xi^{-1  \dagger} = \left ( \begin{array}{cc}
\sqrt{ \frac{\omega_{0}}{\sqrt{\Delta_{1}^2 + v^{2}k^{2}} } } & 0 \\
0 & i\sqrt{ \frac{\omega_{0}}{\sqrt{\Delta_{2}^2 + v^{2}k^{2}} } }
 \end{array} \right ) \end{eqnarray}

While in the limit $\Delta_{2} \rightarrow \Delta_{1}$,
\begin{eqnarray} \chi =\frac{\omega_{0}}{\sqrt{\Delta^2 + v^{2}k^{2}}}\xi
= \sqrt{ \frac{\omega_{0}}
{2\sqrt{\Delta^2 + v^{2}k^{2}} } }
\left ( \begin{array}{cc}
	      1 & -i \\
	       1 & i
		\end{array}  \right ) \end{eqnarray}

It turns out that
\begin{eqnarray}  \vec{\phi}(k,t=0) = \frac{1}{\sqrt{2\omega_{0}}}
[\chi^{\dagger}
\vec{b}^{\dagger}_{-k} + \chi^{T} \vec{b}_{k}] \end{eqnarray}
\begin{eqnarray} \vec{\pi} (k,t=0) = i\sqrt{\frac{\omega_{0}}{2}}
[\xi^{\dagger}
\vec{b}^{\dagger}_{-k} - \xi^{T} \vec{b}_{k}] \end{eqnarray}
Where the $b$'s are the operators which diagonalize $H$.

So now we have all the equipment necessary to calculate
$\vec{l}_{a,b}(k,k^{\prime})$.
\begin{eqnarray} \vec{l}_{a,b}(k,k^{\prime}) =
\frac{-i}{2} \left (\xi^{\ast}(k) \vec{\Sigma} \chi
^{T}(k^{\prime})
+ \chi^{\ast}(k) \vec{\Sigma} \xi^T(k^{\prime}) \right )_{a,b} \end{eqnarray}
where we define the cross product matrix with the Levi-Civita symbol by
$\Sigma^i = \epsilon^{ijk}$. The matrices, $\chi$ and $\xi$,
are now three by three with the
inclusion of the trivially diagonal unmixed component (ie. the three
component):
$\chi_{33} = \sqrt{ \frac{\omega_{0}}{\sqrt{\Delta_{3}^2 + v^{2}k^{2}} } }$ and
$\xi_{33} = \sqrt{ \frac{\sqrt{\Delta_{3}^2 + v^{2}k^{2}} }{\omega_{0}}} $.

We would now like to repeat this diagonalization procedure for the Fermion
model.
The  free Hamiltonian is

\begin{eqnarray} \begin{array}{lll}
{\cal H}(x)  =  & &  \\
\frac{1}{2} \left [  \vec{\psi}_L \cdot iv\partial_x \vec{\psi}_L -
\vec{\psi}_R \cdot iv\partial_x \vec{\psi}_R + i\sum_{i=1}^{3} \Delta_i
(\psi_{R,i} \psi_{L,i} - \psi_{L,i}\psi_{R,i}) + i\vec{h} \cdot
(\vec{\psi}_L \times \vec{\psi}_L+\vec{\psi}_R \times \vec{\psi}_R) \right ]
 & &
\end{array}
\end{eqnarray}
Setting $v=1$, we can write
\begin{eqnarray} \vec{\psi}_R = \int_{0}^{\infty} \frac{dk}{2\pi} \left (
e^{-ik(t-x)} \vec{a}_{R,k} + e^{ik(t-x)} \vec{a}_{R,k}^{\dagger}
\right ) \end{eqnarray}
\begin{eqnarray} \vec{\psi}_L = \int_{0}^{\infty} \frac{dk}{2\pi} \left (
e^{-ik(t+x)} \vec{a}_{L,k} + e^{ik(t+x)} \vec{a}_{L,k}^{\dagger} \right )
\end{eqnarray}
\begin{eqnarray}
\{a_k^i, a_{k^\prime}^{j \dagger} \} = 2\pi \delta_{ij} \delta(k-k^\prime)
\end{eqnarray}
the Hamiltonian density in $k$-space becomes
\begin{eqnarray} H_k = \vec{\alpha}_k^{\dagger} M_k \vec{\alpha}_k
\end{eqnarray}
where
\begin{eqnarray} M = \left ( \begin{array}{cc}
\bf{I} k - i\vec{h} \times & i\bf{\Delta} \\
-i\bf{\Delta} & -\bf{I} k - i\vec{h} \times
\end{array} \right ) \end{eqnarray}
\begin{eqnarray} \vec{\alpha}_k = \left ( \begin{array}{c}
  \vec{a}_{R,k} \\
       \vec{a}_{L,k}^{\dagger} \end{array} \right ) \end{eqnarray}
The idea now is to diagonalize this matrix and find the eigenvalues and
eigenvectors. In other words, find the unitary transformation which
diagonalizes $H$.
 We assume the field is in a direction of
symmetry so that we need only diagonalize a $4 \times 4$ matrix. Given that
the field is in the $3$ direction, the eigenvalues for the mixed states
are $\pm \omega_{\pm}$ from eqn. \ref{enerf}.
It may be more illuminating to write out $M$ in a basis that is more natural
to the $U(1)$ problem. Using

\begin{eqnarray} \left( \begin{array}{c} a^1_{R,k} \\ a^2_{R,k} \end{array}
\right ) = \frac{1}{\sqrt{2}} \left ( \begin{array}{cc}
1 & 1 \\
i & -i \end{array} \right ) \left ( \begin{array}{c}
a^+_{R,k} \\
a^-_{R,k} \end{array} \right ) \end{eqnarray}

\begin{eqnarray} \left ( \begin{array}{c} a^{1 \; \dagger}_{L,k} \\
a^{2 \; \dagger}_{L,k} \end{array} \right )
= \frac{1}{\sqrt{2}} \left ( \begin{array}{cc}
1 & 1 \\
-i & i \end{array} \right ) \left ( \begin{array}{c}
a^{+ \; \dagger}_{L,k} \\
a^{+ \; \dagger}_{L,k}  \end{array} \right ) \end{eqnarray}

In this basis, $M$ becomes
\begin{eqnarray} M = \left ( \begin{array}{cc}
k \bf{I}  - h\bf{\sigma}_3 & i\Delta \bf{\sigma}_1 + i\delta \bf{I} \\
-i\Delta \bf{\sigma}_1-i\delta \bf{I} & -k \bf{I} + h\bf{\sigma}_3
\end{array} \right ) \end{eqnarray}
Where $\Delta = \frac{\Delta_1 + \Delta_2}{2}$ and $\delta = \frac{\Delta_1 -
\Delta_2}{2}$.
The equations for the components of the eigenvectors possess the symmetries
\begin{eqnarray} u_1 \leftrightarrow u_2,  u_3 \leftrightarrow u_4, h
\leftrightarrow -h \end{eqnarray}

\begin{eqnarray} u_1 \leftrightarrow u_3,  u_2 \leftrightarrow u_4, \omega
\leftrightarrow
-\omega \end{eqnarray}
where $\omega$ is the eigenvalue.
After some algebra,
\begin{eqnarray} u_4 = \frac{2i\Delta (k-\omega)}{\omega^2 + \Delta^2 -(k+h)^2
- \delta^2}u_1
\end{eqnarray}
\begin{eqnarray} u_3 = \frac{2i\delta (k-\omega)}{(\omega-h)^2 - k^2 + \delta^2
-\Delta^2}u_1
\end{eqnarray}
\begin{eqnarray} u_2 = \frac{2i\Delta (k+\omega)}{\omega^2 + \Delta^2 -(k+h)^2
- \delta^2}u_3
\end{eqnarray}
Using the normalization condition,
\begin{eqnarray} \sum_{i=1}^{4} |u_i|^2 = 1 \end{eqnarray}
we set the phase of $u_1$ to be real for positive eigenvalues; the above
symmetries allow us
the freedom to choose a convenient phase for $u_1$ corresponding to negative
eigenvalues.
\begin{eqnarray} u_1 & = &  2\delta (k+\omega)
(\omega^2 + \Delta^2 -(k+h)^2 - \delta^2) \div \\ &  &
 [  4\delta^2 (k+\omega)^2 (  (\omega^2 + \Delta^2 -(k+h)^2 - \delta^2)^2 +
 4\Delta^2 (k-\omega)^2 ) + \nonumber \\ & &
 ((\omega+h)^2 - k^2 + \delta^2 -\Delta^2)^2 (
 (\omega^2 + \Delta^2 -(k+h)^2 - \delta^2)^2 + 4\Delta^2 (k+\omega)^2 )
]^\frac{1}{2}
 \nonumber
  \end{eqnarray}

We define the $6\times 6$ diagonalizing matrix with columns given by the
eigenvectors
$\vec{u}_{\omega}$ as
\begin{eqnarray} X_{i,\omega} = ( u_{\omega_+}^i, u_{\omega_-}^i,
u_{\omega_3}^i, u_{-\omega_+
}^i,
u_{-\omega_-}^i, u_{-\omega_3}^i, ) \end{eqnarray}

\begin{eqnarray} \alpha^i_k = U X^i_{\omega} \beta^{\omega} \end{eqnarray}

\begin{eqnarray}
U = \frac{1}{\sqrt{2}} \left ( \begin{array}{cccccc}
1 & 1 & 0 & 0 & 0 & 0 \\
i & -i & 0 & 0 & 0 & 0 \\
0 & 0 & \sqrt{2} & 0 & 0 & 0 \\
0 & 0 & 0 & 1 & 1 & 0 \\
0 & 0 & 0 & -i & i & 0 \\
0 & 0 & 0 & 0 & 0 & \sqrt{2}
\end{array} \right ) \equiv \left ( \begin{array}{cc}
V & 0 \\
0 & V \sigma_1  \end{array} \right )
\end{eqnarray}

The diagonal operators, $\beta^{\omega}$ are defined as:
\begin{eqnarray} \vec{\beta}_k = \left ( \begin{array}{c}
  \vec{c}_k \\
	 \vec{d}_{k}^{\dagger} \end{array} \right ) \end{eqnarray}

Our freedom in choosing the phases for the eigenvectors corresponding to
negative
eigenvalues allow us to write

\begin{eqnarray}
X = \left ( \begin{array}{cc}
R & T \\
T & R \end{array} \right )
\end{eqnarray}

The $d$'s and $c$'s correspond to left and right moving fermions, respectively.
This becomes clear in the limit $\Delta_1 \rightarrow \Delta_2 \rightarrow 0$.

Some limiting forms of $R$ and $T$ are:

\begin{eqnarray}
R(h \rightarrow 0 ) = \frac{1}{2} \left ( \begin{array}{ccc}
\sqrt{\frac{\omega_1+k}{\omega_1}} & -\sqrt{\frac{\omega_2+k}{\omega_2}} & 0 \\
\sqrt{\frac{\omega_1+k}{\omega_1}} & \sqrt{\frac{\omega_2+k}{\omega_2}} & 0 \\
0 & 0 & \sqrt{2\frac{\omega_3+k}{\omega_3}}
\end{array} \right )
\end{eqnarray}

\begin{eqnarray}
T(h \rightarrow 0 ) = \frac{1}{2} \left ( \begin{array}{ccc}
-i\sqrt{\frac{\omega_1-k}{\omega_1}} & -i\sqrt{\frac{\omega_2-k}{\omega_2}} & 0
\\
-i\sqrt{\frac{\omega_1-k}{\omega_1}} & i\sqrt{\frac{\omega_2-k}{\omega_2}} & 0
\\
0 & 0 & -i\sqrt{2\frac{\omega_3-k}{\omega_3}}
\end{array} \right )
\end{eqnarray}

\begin{eqnarray}
R(\delta \rightarrow 0 ) = \frac{1}{\sqrt{2}} \left ( \begin{array}{ccc}
0 & -\sqrt{\frac{\omega_{\Delta}+k}{\omega_{\Delta}}} & 0 \\
\sqrt{\frac{\omega_{\Delta}+k}{\omega_{\Delta}}} & 0 & 0 \\
0 & 0 & \sqrt{\frac{\omega_3+k}{\omega_3}}
\end{array} \right )
\end{eqnarray}

\begin{eqnarray}
T(\delta \rightarrow 0 ) = \frac{1}{\sqrt{2}} \left ( \begin{array}{ccc}
-i\sqrt{\frac{\omega_{\Delta}-k}{\omega_{\Delta}}} & 0 & 0 \\
0 & i\sqrt{\frac{\omega_{\Delta}-k}{\omega_{\Delta}}} & 0 \\
0 & 0 & -i\sqrt{\frac{\omega_3-k}{\omega_3}}
\end{array} \right )
\end{eqnarray}

The expression for the uniform part of the spin in terms of fermions is
\begin{eqnarray} \frac{-i}{2} \left ( \vec{\psi_L} \times
\vec{\psi_L}+\vec{\psi_R} \times \vec{\psi_R} \right ) \end{eqnarray}
We can write $\vec{l}_{a,b}(k,k^{\prime})$  as
\begin{eqnarray}  \vec{l}_{a,b}(k,k^{\prime}) =-i\int_{0}^{\infty} dp \; dq \;
<k,a|(\vec{\alpha}^{\dagger}_p + \vec{\alpha}^{T}_p)
\left ( \begin{array}{cc} \vec{\Sigma} & 0 \\
0 & \vec{\Sigma} \end{array} \right ) (\vec{\alpha}_q + \vec{\alpha}^{\ast}_q)
|k^{\prime},b> \end{eqnarray}
where again $\Sigma$ is the cross product matrix.

After some algebra, we get

\begin{eqnarray}
\vec{l}_{a,b}(k,k^{\prime}) =
-i \left ( \begin{array}{cc}
R^{\dagger} V^{\dagger} \vec{\Sigma}VR + T^{\dagger} \sigma_1 V^{\dagger}
\vec{\Sigma}
V \sigma_1 T & R^{\dagger} V^{\dagger} \vec{\Sigma} V^{\ast} T^{\ast} +
T^{\dagger} \sigma_1 V^{\dagger} \vec{\Sigma} V^{\ast} \sigma_1 R^{\ast} \\
T^T V^T \vec{\Sigma} VR + R^T \sigma_1 V^T \vec{\Sigma} V \sigma_1 T &
T^T V^T \vec{\Sigma}V^{\ast} T^{\ast} + R^T \sigma_1 V^T \vec{\Sigma}
V^{\ast} \sigma_1 R^{\ast} \end{array} \right )
\end{eqnarray}
Each index of this matrix runs over six states; the first and last three
correspond to right and left movers respectively. In the case of $U(1)$
symmetry each set would correspond to states of definite spin.

\section{Impurity Effects on $1/T_1$}
\label{appB}

The NMR experiments done on NENP have exhibited curious behaviour at small
fields
(0-4 T) and low temperatures. In this regime, our calculations have shown that
$1/T_1$ should decrease rapidly with decreasing applied field. Experiments,
however, indicate
a non-zero lifetime at $h=0$ that undergoes a minimum at about 2 T before
showing the
expected exponential behaviour at around 4 T. Moreover, the zero field lifetime
increases with
temperature.\cite{fuji}

If one attempted to fit the mid field data (which is expected to be in good
agreement with our calculations) to our results and assumed that there is no
other mechanism for relaxation, then the observed values would deviate from
the calculated ones at low fields ($h = .2-3$ Tesla) by a factor as big as
300 for $T=1.4$ Kelvin.

Clearly the low field bump is an impurity effect. Dynamics due to dimensional
crossover are ruled out by the fact that $1/T_1$ increases with temperature at
the low fields discussed. A possible, yet not fully convincing contribution can
come from the dynamics
of the spin-$\frac{1}{2}$ degrees of freedom at the ends of large but {\em
finite}
chains in NENP. This is discussed further in ref. [\onlinecite{mit}].
The essential
physics is that the resonance energy of a spin-$\frac{1}{2}$ at a chain end
in a uniform field crossed with an RF field
is broadened by the interaction with thermally excited magnons even at
temperatures
well below the gap. At the Larmour frequency of a Hydrogen nucleus and low
magnetic field, the Fourier
transform of the local spin-spin correlator (which is proportional to $1/T_1$)
is of the form:\cite{mit}

\begin{eqnarray}
1/T_{1_{Imp}} \propto e^{-\Delta/T}
\end{eqnarray}
This is true in both the limits of long and short chains. In both instances
this is not consistent with the weaker temperature dependence seen at constant
low magnetic field. We conclude that the experimental results are due to
a mechanism not intrinsic to the Haldane system or coupled to it.

\section{The Structure of NENP and Experimental Ramifications}
\label{appC}

Each chain in NENP is comprised of Ethylenediamine-Nickel chelates separated by
nitrite groups. It is important to realize, however,
that two neighbouring $Ni^{2+}$ are not equivalent; rather, one is related to
the other by a $\pi$ rotation about the $b$ axis. Also, the angle
along the $N-Ni-O$ bond is not exactly $\pi$, meaning that the $Ni$ site is not
truly
centrosymmetric. Most importantly, the local symmetry axes of each $Ni$ ion are
rotated with respect to the $abc$ (crystallographic) axes. To demonstrate this
we now note the
coordinates of the Nitrogen atoms in the Ethylenediamine chelate surrounding
the Nickel (placing the Nickel at
the origin): \cite{mey}

\begin{center}
\begin{tabular}{|c|c|c|c|}
 Atom  &  $a$ $(\AA)$ & $b$ $(\AA)$ & $c$ $(\AA)$  \\ \hline
 $N(1)$ &  $2.053$ $(3)$  &  $.162$ $(3)$  &  $.338$ $(3)$ \\
     &   &   &   \\
     $N(2)$ &  $.619$ $(3)$  &  $-.184$ $(3)$  &  $-1.971$ $(3)$ \\
       &   &   &   \\
       \end{tabular}
\end{center}

\begin{figure}
\epsfxsize=14 cm
\epsffile{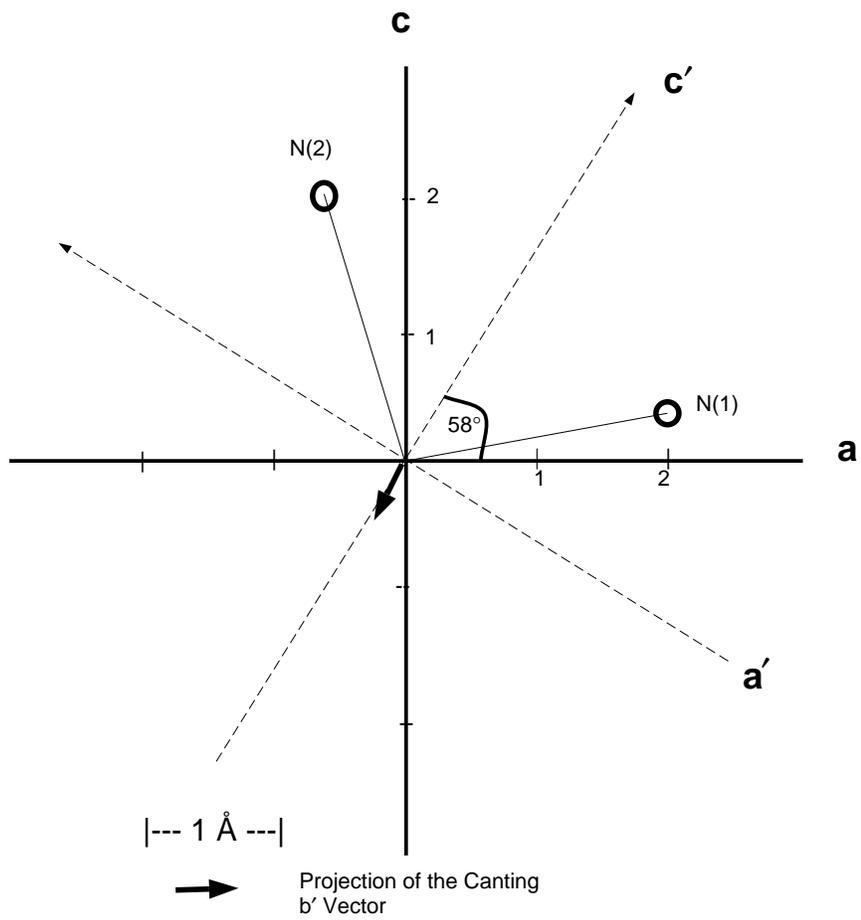}
\caption{Local and crystallographic axes projected onto the $ac$-plane in NENP}
\label{fig:7}
\end{figure}

The other Nitrogen atoms in the chelate can be obtained by reflection
through the Nickel. One easily sees that projecting this structure onto
the $b$ plane yields symmetry axes (in the $b$ plane) rotated $\sim 60^{\circ}$
from the $ac$-axes. This is shown in Fig. \ref{fig:7}. The inclination
of the local Nickel axes from the $abc$ system can be obtained by taking
the cross product of the two Nitrogen vectors (ie. the normal to the plane
described by the four Nitrogen atoms in the chelate:
\begin{eqnarray}
\hat{n} = \left ( -.06 \mbox{ } (1) ,.98 \mbox{ } (1) ,-.11 \mbox{ } (1)
\right )
\end{eqnarray}

{}From the above, it is clear that the local $Ni$ $b^{\prime}$-axis makes a
$\sim 10^{\circ}$
angle with the $b$-axis, while the azimuthal angle in the $ac$ plane is
$\sim -28^{\circ}$ from $c$.
The $10^{\circ}$ tilt is roughly about the $a^{\prime}$ direction of the local
symmetry axes.

One may worry that the $NO_2$ radicals may distort the local symmetry axes, but
remarkably enough,
when projected onto the $ac$ plane, the three atoms in the molecule sit on the
$c^\prime$ axis. This
reinforces our suspicion that the local symmetry axes are indeed what we have
above.

\begin{figure}
\epsfxsize=14 cm
\epsffile{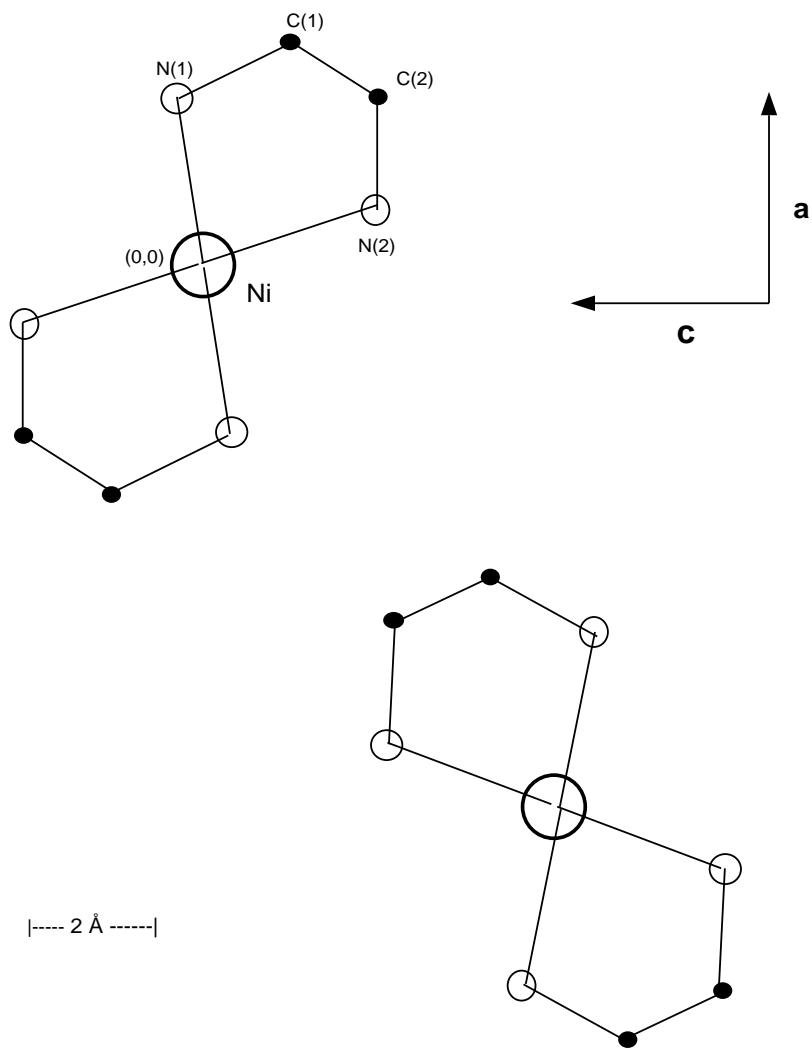}
\caption{A projection of the NENP unit cell onto the $ac$-plane showing two
chains per unit cell}
\label{fig:8}
\end{figure}

Next, we consider the whole space group of NENP. The most recent attempt to
solve for the crystal symmetries has concluded that the true space group of
the material is $Pn2_{1}a$ \cite{mey}; this is a non-centrosymmetric space
group with a
screw $2_{1}$ symmetry about the $b$ axis, diagonal glide plane reflection
symmetry
along the $a$ axis, and an axial glide plane reflection symmetry along $c$.
 Experimentally, attempts to
solve the structure in $Pn2_{1}a$ have not been successful; rather, it seems
that
$Pnma$ gives a better fit. The main difference between the two is the presence
in $Pnma$
of a mirror plane parallel to $b$ at $\frac{1}{4}b$, centers of symmetry at
various locations
in the unit cell, and two-fold screw axes separating these centers of symmetry.
The reason for the experimental discrepancy is attributed to disorder
in the orientation of the nitrite group, the perchlorate anions, and the
existence
of a local or pseudo center of symmetry lying very close to the $Ni$
(thousandths
of an Angstrom) \cite{mey}. A crucial point is that both space groups
share the axial glide planes along $a$, the diagonal glide planes along
$C$ and the $2_{1}$ screw symmetry about $b$. These
generate a total of 4 $Ni$ sites per primitive cell and two chains through
each cell. The two chains are such that the $Ni$ chelates on one are the mirror
image
of the other. Figure \ref{fig:8} shows a projection of this picture onto
the $ac$ plane.

The presence of the $2_{1}$ screw symmetry about each chain axis introduces
staggered
contributions to the local anisotropy and gyromagnetic tensors. This is
because, as motivated above,
these are not diagonal in the crystallographic coordinate system. The resulting
spin
Hamiltonian is
  \begin{eqnarray}
  H = J\sum_{i}  \left [
\vec{S}_{i} \cdot \vec{S}_{i+1}  +
\vec{S}_{i} \cdot {\bf D} \vec{S}_{i} - \mu_{B} \vec{S}_{i} \cdot {\bf G}
\vec{H
} +
(-1)^{i} (\vec{S}_{i} \cdot {\bf d} \vec{S}_{i} - \mu_{B} \vec{S}_{i} \cdot
{\bf
 g} \vec{H})
 \right ]
 \end{eqnarray}

We make the assumption that the symmetry of the mass and $g$-tensors
is the same (ie. that at each site they can be simultaneously diagonalized).
We can get the required parametrization for the $g$-tensors from high
temperature uniform susceptibility data \cite{mey}. This is based on the idea
that at
high temperatures the $Ni$ atoms will behave as an ensemble of uncoupled spins
($s=1$)
with the same gyromagnetic tensor as in the antiferromagnetic case.
With this in mind we get

  \begin{eqnarray}
  \bf{G} = \left ( \begin{array}{ccc}
G_{c^{\prime}} \cos^{2} (\theta) + G_{b^{\prime}} \sin^{2} (\theta) & 0 & 0 \\
	       0 & G_{a^{\prime}} & 0   \\
		    0 & 0 & G_{b^{\prime}} \cos^{2} (\theta) + G_{c^{\prime}} \sin^{2}
(\theta)
			 \end{array} \right )
\end{eqnarray}

  \begin{eqnarray}
  \bf{g} = \left ( \begin{array}{ccc}
		     0 & 0 & \sin(\theta) \cos (\theta) (G_{b^{\prime}} - G_{c^{\prime}}) \\
		     0 & 0 & 0 \\
		     \sin(\theta) \cos (\theta) (G_{b^{\prime}} - G_{c^{\prime}}) & 0 & 0
                         \end{array} \right )
\end{eqnarray}

Here $\theta \sim 10^\circ $, and $G_{a^{\prime}} = 2.24$, $G_{b^{\prime}} =
2.15$,
$G_{c^{\prime}} = 2.20$
are the values for the local $G$-tensor that give the observed high temperature
$g$-tensor when averaged over the unit cell.
Correspondingly, the mass tensors must have the following form:
 \begin{eqnarray}
\bf{D} = \left ( \begin{array}{cccc}
D_{c^{\prime}} & 0 & 0 \\
0 & D_{a^{\prime}} & 0 \\
0 & 0 & D_{b^{\prime}}
\end{array} \right )
\end{eqnarray}

  \begin{eqnarray}
  \bf{d} = \left ( \begin{array}{cccc}
	    0 & 0 & \frac{\tan(2\theta)}{2} (D_{b^{\prime}} - D_{c^{\prime}}) \\
                  0 & 0 & 0 \\
				\frac{\tan(2\theta)}{2} (D_{b^{\prime}} - D_{c^{\prime}}) & 0 & 0
                 \end{array} \right )
\end{eqnarray}

The parameters $D_{a^{\prime}},D_{c^{\prime}},D_{b^{\prime}}$
are to be fitted by experiment to the model used to describe
the system.

Using
 \begin{eqnarray}
 \vec{S_i} = (-1)^i \vec{\phi} + \vec{\phi} \times \vec{\Pi}
 \end{eqnarray}
The boson Hamiltonian can now be written

  \begin{eqnarray}
  H = \int dx  [ \frac{v}{2} \vec{\Pi}^2 + \frac{v}{2} (\frac{\partial
 \vec{\phi} }{\partial x})^2 + \frac{1}{2v} \vec{\phi} \cdot {\bf D}
 \vec{\phi} - \mu_{B} \vec{H} \cdot {\bf G} (\vec{\phi} \times \vec{\Pi}) + \\
 \frac{1}{v} \vec{\phi} \cdot {\bf d} (\vec{\phi} \times \vec{\Pi})
  - \mu_{B} \vec{\phi} \cdot {\bf g} \vec{H} + \lambda (\vec{\phi}^2)^2  ]
\nonumber
   \end{eqnarray}

The term containing {\bf d} breaks the $Z_2$ symmetry along the $a^{\prime}$
(lowest mass)
direction. It will also renormalize the masses. The second effect can be
ignored
in the approximation that the $\phi^4$ term is ignored if we assume the masses
are
physical. The first effect leads to the presence of a static staggered field
even
below a critical magnetic field. A gap will always persist. The staggered field
term
will break the $Z_2$ symmetry along the $c^{\prime}$ or $b$ axis, depending on
whether the field is
applied in the $b$ or $c^{\prime}$ direction, respectively. A static staggered
moment will
likewise appear due to this term.

We would now like to discuss the effect of having two inequivalent chains per
unit cell,
with local axes different from the crystalographic axes.
We label the two chains found in a unit cell of NENP
`chain 1' and `chain 2' corresponding to the chains in the upper right and
lower left corners of Figure \ref{fig:8} respectively. The dispersion
branches of chain 1 are given by (20) of [\onlinecite{aff}] only the field is
$\alpha - 30^\circ$ from the $c^{\prime}$ axis where $\alpha$ is the angle
of the field from the crystallographic $c$-axis. Similarly, the dispersion
branches of chain 2 are calculated with the field $\alpha -150^\circ$
from the $c^{\prime}$ axis. We can now graph these and compare
with experiment.

Experiments which average over signals, like susceptibility or NMR $T_1^{-1}$
measurements, must consider their results an average of two different
experiments
(corresponding to the two different chains with their relatively different
field
configurations). on the other hand, experiments such as ESR, should show a
separate
signal for each chain. The NMR relaxation calculations performed in the main
part of the
text ignore these effects. This ought not make a difference to the
qualitative conclusions.

\begin{figure}
\epsfxsize=14 cm
\epsffile{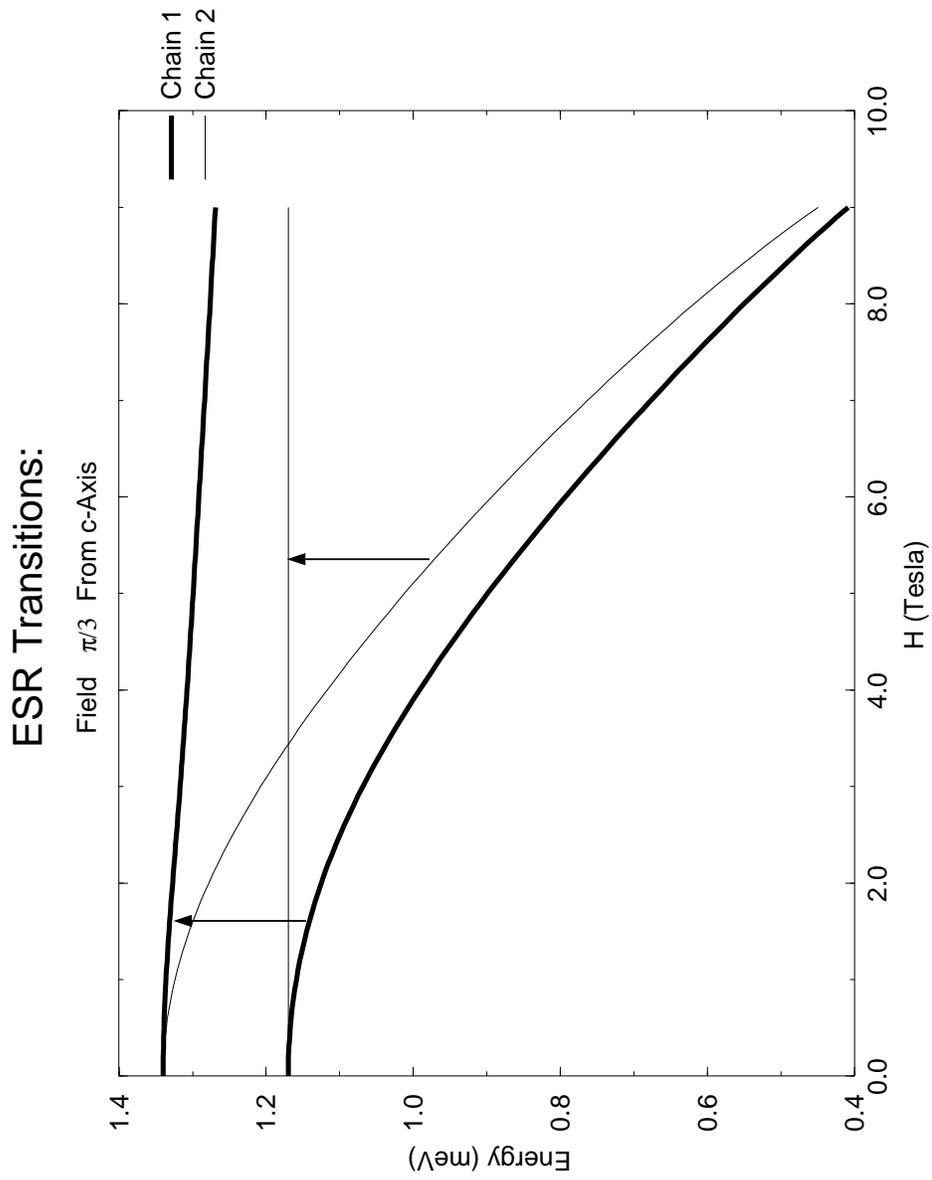}
\caption{Dispersions for the two chain conformations and sample resonant
transitions
for a uniform field placed $60^\circ$ from the crystallographic $c$-axis.}
\label{fig:9}
\end{figure}

Figure \ref{fig:9}  shows the dispersions for chains 1 and 2 (solid and dashed
lines, respectively) when the field is $\pi /3$ from the crystallographic $c$
axis in the $ac$-plane.
This is an example of how transitions at two
field strengths ought to be possible in the ESR experiment.

\begin{figure}
\epsfxsize=14 cm
\epsffile{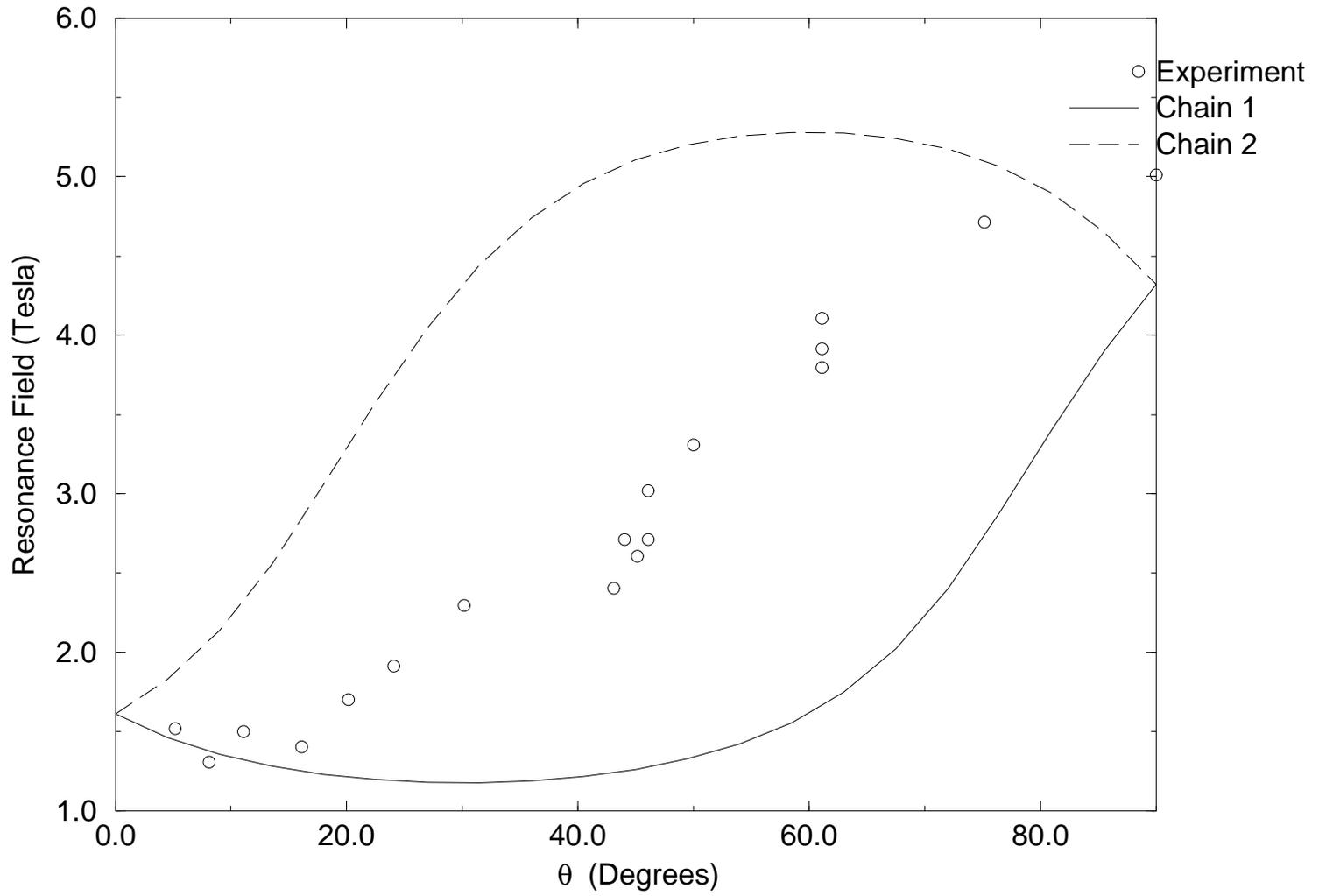}
\caption{Resonant field vs. field orientation in the $ac$-plane for .19 meV
transitions.}
\label{fig:10}
\end{figure}

Figure \ref{fig:10} shows the resonance field versus orientation of field in
the crystallographic $ac$-plane. The lower branch denotes transitions in chain
1
while the upper branch corresponds to transitions in chain 2.
The transitions were
calculated at .19 meV.
corresponding to 47 GHz
In addition the experimental
results of Date and Kindo \cite{dat} are represented by the crosses. One
immediately
sees that the data does not compare well with the predictions based on the
free model used above, for instead of
following one of the branches, the experimental results lie between them.
Furthermore, it seems unlikely that perturbations will cause such a significant
shift in the resonance field. One sees that the discrepancy is $\sim
\pm 1$ Tesla. One possible explanation is that since the ESR signal
in [\onlinecite{dat}]
was also $\sim \pm 1$ Tesla in width and symmetric (in conflict with the
predictions of [\onlinecite{aff})], the signal from the resonances in both
chains was somehow smeared and interpreted as one single peak. Seen that way
the model predictions are in good agreement except for the large field
regime. One also has to keep in mind that the high-field boson dispersions
are not accurate and therefore the predictions at larger angles could easily be
.5 Tesla (or more) off the mark.

To end this discussion on, we'd like to elaborate on
a previously made statement regarding the assignment of masses to the
local Ni symmetry axes. It's easy to see that switching the masses around
is tantamount to a $\pi/2$ shift in Figure \ref{fig:7}
(the fact that the gyromagnetic
constants are not the same in orthogonal directions will not change the
ESR resonance graph much since the ratio of the gyromagnetic constants is
0.98).
Redrawing Figure \ref{fig:10} with this geometry misses the experimental
results by 4 Tesla at 0 and 90 degrees,
where the two chain resonances coincide.
This determines the proper labeling of the local symmetry axes.

\begin{figure}
\epsfxsize=14 cm
\epsffile{fig5}
\caption{F(h,T) for fermions: field along crystal $b$ and $c$ directions}
\label{fig:5}
\end{figure}

\begin{thebibliography}{99}

\bibitem{hal1} F.D.M. Haldane, Phys. Lett. {\bf 93A}, 464 (1983). For
a review see I. Affleck, J. Phys.:Condensed Matter {\bf 1}, 2047 (1989).

\bibitem{aff1} I. Affleck, Phys. Rev. B {\bf 41}, 6697 (1990).

\bibitem{tsev} A.M. Tsvelik, Phys. Rev. B {\bf 42}, 10499 (1990).

\bibitem{Aff} I. Affleck, Phys. Rev. B {\bf 43}, 3215 (1991).

\bibitem{wes} I. Affleck and R.A. Weston, Phys. Rev. B {\bf 45}, 4667 (1992).

\bibitem{mit} P. P. Mitra, B.I. Halperin and I. Affleck, Phys. Rev. B {\bf 45},
5299 (1992).

\bibitem{aff} I. Affleck, Phys. Rev. B {\bf 46}, 9002 (1992).

\bibitem{joli} Th. Jolicoeur and O. Golinelli,  Phys. Rev. B {\bf 50}, 9265,
(1994).

\bibitem{fuji} N. Fujiwara et. al., Phys. Rev. B {\bf 47}, 11860 (1993).

\bibitem{aff2} I. Affleck, Phys. Rev. Lett. {\bf 56}, 746 (1986).

\bibitem{sor} E. S. S$\mbox{\o}$renson and I. Affleck, Phys. Rev. B {\bf 49},
13235 (1994) and Phys. Rev. B {\bf 49}, 15771 (1994).

\bibitem{halp} P. P. Mitra and  B.I. Halperin, Preprint.

\bibitem{sachdev} S. Sachdev et. al.,  Phys. Rev. B {\bf 50}, 13006, 1994.

\bibitem{troy} M. Troyer et. al., Phys. Rev. B {\bf 50}, 13515 (1994).

\bibitem{mah} G.D. Mahan, {\em Many Particle Physics} (Plenum Press, 1993),
p. 604.

\bibitem{taki} M. Takigawa et. al., Preprint

\bibitem{car} J.L. Cardy, Nuc. Phys. B{\bf 270}, 186 (1986).

\bibitem{bla} Jean-Paul Blaizot and Georges Ripka, {\em Quantum Theory
of Finite Systems} (MIT Press, 1986), Chapter 3.

\bibitem{mey} A. Meyer et. al., Inorg. Chem. {\bf 21}, 1729 (1982).

\bibitem{dat} M. Date and K. Kindo, Phys. Rev. Lett. {\bf 65}, 1659 (1990).


\end{thebibliography}
\end{document}